\documentclass[12pt, notitlepage]{article}
\pdfoutput=1

\usepackage{float}
\usepackage{subfig}
\usepackage{graphicx}
\usepackage[T1]{fontenc}
\usepackage[utf8]{inputenc}
\usepackage{authblk}
\usepackage{amsmath,bm}
\usepackage[left=2cm,right=2cm,top=2cm,bottom=2cm]{geometry}

\renewcommand\footnotemark{}
\usepackage[dvipsnames]{xcolor}

\newcommand{\rf}[1]{(\ref{#1})}
\newcommand{\beq}{\begin{equation}}
\newcommand{\beql}[1]{\beq\label{#1}}
\newcommand{\eeq}{\end{equation}}
\newcommand{\bea}{\begin{eqnarray}}
\newcommand{\eea}{\end{eqnarray}}

\begin{document}

\title{The higher-order phase transition in toroidal CDT}

\author[a,b]{J.~Ambj\o rn}
\author[c]{G.~Czelusta}
\author[c]{J.~Gizbert-Studnicki}
\author[c]{A.~G\"orlich}
\author[c]{J.~Jurkiewicz}
\author[c]{D.~N\'emeth}
\affil[a]{\small{The Niels Bohr Institute, Copenhagen University, \authorcr Blegdamsvej 17, DK-2100 Copenhagen Ø, Denmark. \authorcr E-mail: ambjorn@nbi.dk.\vspace{+2ex}}} 

\affil[b]{\small{IMAPP, Radboud University, \authorcr Nijmegen, PO Box 9010, The Netherlands.\vspace{+2ex}}}

\affil[c]{\small{Institute of Theoretical Physics, Jagiellonian University, \authorcr \L ojasiewicza 11, Krak\'ow, PL 30-348, Poland. \authorcr Email: grzegorz.czelusta@doctoral.uj.edu.pl, jakub.gizbert-studnicki@uj.edu.pl, andrzej.goerlich@uj.edu.pl, jerzy.jurkiewicz@uj.edu.pl, nemeth.daniel.1992@gmail.com.}}

\date{\small({Dated: \today})}          
\maketitle


\begin{abstract}

We investigate the  transition between the phases $B$ and $C_b$ 
observed in  four-dimensional Causal Dynamical Triangulations (CDT). We find that the critical properties of CDT with toroidal spatial topology are the same as earlier observed in spherical spatial topology where the $B-C_b$ transition was found to be higher-order. This  may have important consequences for the existence of the  continuum limit of CDT, describing the perspective UV limit of {\it quantum gravity}, which potentially can be investigated in the toroidal model.

\vspace{1cm}

\end{abstract}

\begin{section}{Introduction}\label{intro}

Numerical Monte Carlo simulations  applied to  lattice field theories became an important tool of contemporary  physics. The famous example is Lattice Quantum Chromodynamics (QCD) which has grown up from  its  childhood and now goes hand-by-hand with  experiments and beyond, e.g. by investigating  very non-trivial QCD phase diagram in the regime of coupling constants non-tractable by  perturbative calculus. Despite many open questions, QCD has a well defined ultraviolet limit, where it becomes non-interacting {\it asymptotically free} theory and thus the high energy behaviour can be investigated perturbatively. The opposite thing happens when one tries to formulate a quantum theory of gravity (QG) by applying standard quantum field theory techniques to Einstein's General Relativity (GR). In that case the perturbative expansion around any fixed classical metric field fails at high energies due to the perturbative non-renormalizability of such a formulation \cite{'tHooft:1974bx,Goroff:1985th}.  However, as conjectured by Steven Weinberg in his seminal paper \cite{Weinberg79},  QG can be {\it asymptotically safe}, i.e. it can admit a  well behaved non-perturbative high energy limit defined  in the vicinity of a non-trivial fixed point of the renormalization group flow,  where quantum gravity becomes scale-invariant and thus can be extrapolated to arbitrarily large energy scale. If the {\it asymptotic safety} scenario is valid\footnote{There is growing evidence for the existence of a fixed point suitable for asymptotic safety coming from functional renormalization group studies \cite{Reuter:1998,Litim:2003vp, Niedermaier2006, Codello:2008vh,Benedetti:2009rx,Litim:2011cp}, however a rigorous proof of its existence is still lacking.} then (in the ultraviolet regime) QG must be formulated in a background-independent non-perturbative way making lattice approaches well suited  to tackle this problem. In such formulations one discretizes geometric degrees of freedom on the lattice with (4-dimensional) lattice 'volume' $N_4$  and with a minimal (cut-off) spacing $a$, and in the ultraviolet regime one would like to get rid of the discretization by taking a continuum limit of $a\to0$ and $N_4\to\infty$ such that $N_4^{1/4} \cdot a$ is related to some physical length. In order to obtain  non-trivial  physical observables  in the continuum limit, where $a\to0$ and $N_4\to\infty$, one would also like to have  appropriately divergent correlation lengths $\ell_c \sim N_4^{1/4}$. Thus in a lattice approach the continuum limit should be associated with a higher order (continuous) phase transition. Therefore studies of the phase structure  and orders of phase transitions are  important steps towards defining an ultraviolet limit in a lattice formulation and thus testing  the asymptotic safety scenario for gravity.
\newline

One of the most successful attempts of the lattice formulation of  quantum gravity is that of Causal Dynamical
Triangulations (CDT) (for reviews see \cite{Ambjorn:2012jv,Loll:2019rdj}), in the sense that it has a rich phase structure, where some of the transitions are 
higher order, 
which potentially can be used to define continuum limit and that it additionally has a well behaved low energy limit 
consistent with GR. 
 CDT is based on the path integral formalism and makes only a few assumptions on the geometry of quantum space-time, namely it requires that the geometry can be globally foliated into space-like hypersurfaces, each with the same fixed topology $\Sigma$.
 The model is using the discretization of space-time following the method proposed by Regge \cite{Regge:1961px}. 
 The three-dimensional
 spacial states are constructed by gluing together in all possible ways regular tetrahedra with a common link length
 $a_s$ to form a triangulation of a three-dimensional space with a (closed) topology $\Sigma$. The topology of
 states is fixed during the evolution of geometry in time, being the origin of the name {\it causality} in the model.
 To join states at different times $t$ we need two types of 4-dimensional simplices.
 Tetrahedra become bases of 4-dimensional simplices $\{4,1\}$ (and $\{1,4\}$) 
 with four vertices at a time layer $t$ and one at $t+1$ (resp. $t-1$).  In our notation the simplex $\{i,j\}$ 
 has $i$ vertices at a time $t$ and $j$ vertices at a time $t+1$.
 The time links are assumed to have a common link length $a_t$ which may be
 different than $a_s$. To complete the manifold structure two additional simplex structures are necessary.
 These are $\{3,2\}$ and $\{2,3\}$ simplices. Pairs of simplices share a common
 three-dimensional face (tetrahedron). The construction works both for systems with Lorentzian signature and,
 after Wick rotation, for systems with Euclidean signature. 
 Each space-time configuration can be
 interpreted as Lorentzian or Euclidean. The possibility of performing Wick rotation is crucial if we want
 to use numerical methods to analyze the properties of the model. In the following, we assume the Euclidean
 formulation is used. The discretization described above means that the four-dimensional volume of all
 $\{i,j\}$ simplices depends only on the type of a simplex. Similarly other geometric properties, like the
 angles, are universal for all simplices of a particular type.
 
 The studied object is the Feynman amplitude ${\cal Z}$, which is expressed as a weighted sum over manifolds ${\cal T}$ joining
 the initial and final geometric states separated by time $T$. The weight is assumed to be expressed as a discretized version of the
 Hilbert-Einstein action $S_{EH}({\cal T})$ 
 
\begin{equation} \label{Feynman}
{\cal Z} =\sum_{\cal T} \frac{1}{C({\cal T})}  e^{-S_{EH}},
\end{equation}
where $C({\cal T})$ is the symmetry factor of a graph representing the manifold. In practice the choice of the 
initial and final states is replaced by assuming the system to be periodic with the period $T$.
The discretized version of the Hilbert-Einstein action takes the form \cite{Ambjorn:2001cv}

 
\begin{equation} \label{bareaction}
S_{EH}=-\left(\kappa_{0}+6\Delta\right)N_{0}+\kappa_{4}\left(N_{4,1}+N_{3,2}\right)+\Delta N_{4,1},
\end{equation}

\noindent where $N_{i,j}$ denotes the number of $4$-dimensional simplicial building blocks with $i$ vertices on hypersurface $t$ and $j$ vertices on hypersurface $t\pm 1$, and $N_0$ is the number of vertices in the triangulation. $\kappa_{0}$, $\Delta$ and $\kappa_{4}$ are bare coupling constants. $\kappa_{0}$ and $\kappa_{4}$ are related to Newton's constant and the cosmological constant, respectively, and $\Delta$ depends on the ratio of the length of space-like and time-like links in the lattice. In the Monte Carlo simulations of CDT the parameter $\kappa_{4}$, which is proportional to the cosmological constant, is tuned such that one can take  infinite-volume limit.
As will be explained later, in numerical simulations we perform a series of measurements for systems with
increasing (fixed) volume $N_{4,1}$ and try to determine the limiting behaviour for $N_{4,1}\to\infty$.
In the consequence the phase diagram presented in Figure \ref{pdnew} depends only on two bare couplings
$\kappa_0$ and $\Delta$. It is remarkable that such a simple model has a rich phase structure with four
phases with very different physical properties. The analysis of the phase structure and, in particular, 
the order of phase transitions is fundamental to relate the model to a possible theory of quantum gravity.

\begin{figure}[H]
\centering
\includegraphics[width=0.53\linewidth]{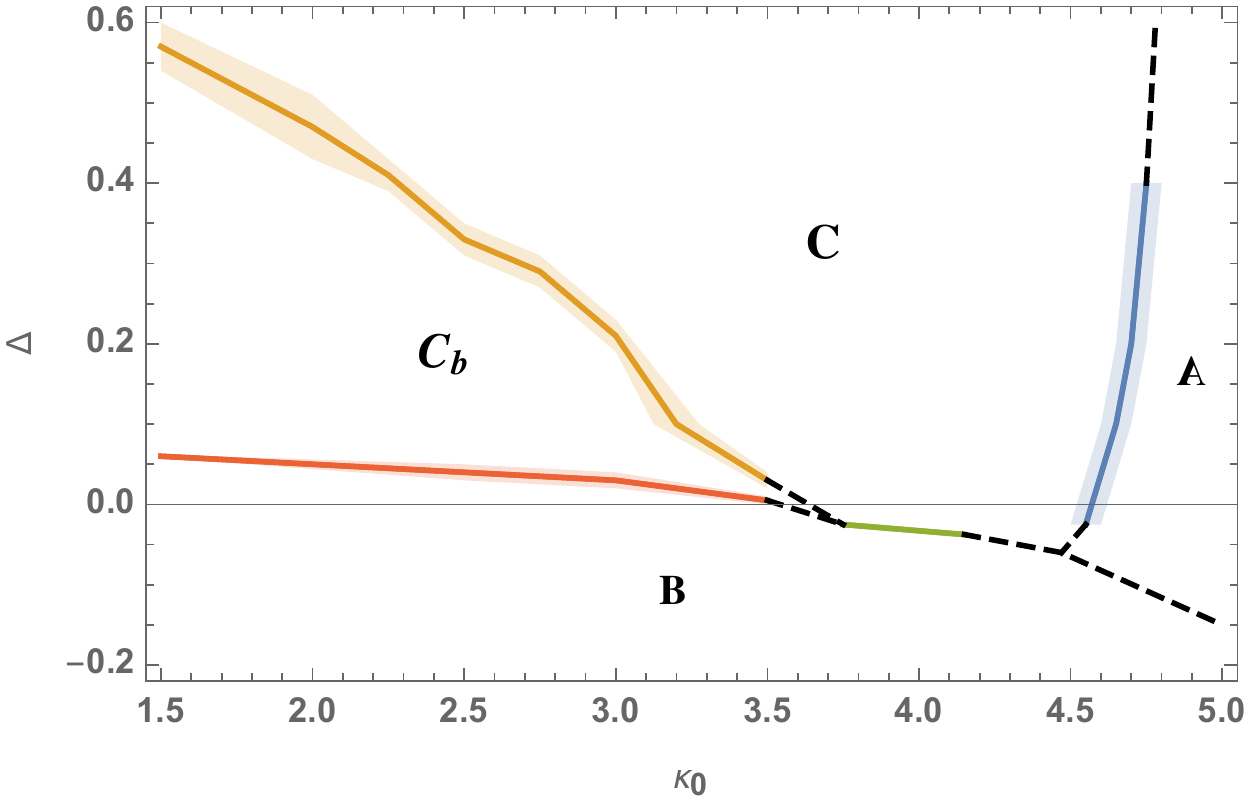}
\caption{\small The phase structure of $4$-dimensional CDT.}
\label{pdnew}
\end{figure}


\end{section}

\begin{section}{Phase transitions  in MC simulations of  lattice field theories}\label{method}

According to  Ehrenfest's classification, the order of a phase transition depends on the behaviour of the thermodynamic free energy. If all first $n-1$ order derivatives of the  free energy  are continuous functions of some thermodynamic variable, e.g. the coupling constant of the lattice theory, and the $n$-th order derivative exhibits a discontinuity  at the transition point   then the transition is the {\it $n$-th order} phase transition. Here we are especially interested to distinguish between the first- and the higher-order phase transitions, as the  continuous limit of the lattice field theory should be associated with the latter type. 

The derivatives of  free energy are related to order parameters, which capture differences of thermodynamic properties of the system in  two different phases separated by the transition point. For a first-order transition one should observe a discontinuity of the order parameter at the transition point and for the higher-order transition the order parameter should be continuous but its derivatives, e.g. its susceptibility, should diverge.
Unfortunately measuring the (dis-)continuity of the (derivatives) of an order parameter in  numerical simulations  is a tedious task. Actually, in numerical Monte Carlo simulations, which are always performed for a finite lattice size $N_4$, one does not even observe phase transitions per se. The finite lattice size and the finite lattice spacing make all thermodynamic functions and their derivatives finite, even though they can become arbitrarily large for large lattice sizes. One should therefore carefully analyze finite (lattice) size effects and, if possible, take the infinite (lattice) volume  limit $N_4\to\infty$.

As phase transitions are usually related to breaking some symmetries of the studied lattice field theory, one can define order parameter(s) $OP$ which capture these symmetry differences between various phases of the theory in question. One then usually performs numerical Monte Carlo (MC) simulations for some fixed lattice volume $N_4$ in many points of the theory parameter space (see e.g. the CDT phase diagram in Figure \ref{pdnew}) to find regions where the order parameter rapidly changes, see e.g. Figure \ref{OPmean} where we show the mean value $\langle OP \rangle$ of the four order parameters (for their definitions see equation \rf{OPs})  used in CDT phase transition studies  measured 
in the $B-C_b$ transition region. The precise position of the phase transition is signaled by a peak of the susceptibility of an order parameter
\beql{susc}
\chi_{OP} \equiv \langle OP^2 \rangle - \langle OP \rangle^2 \ 
\eeq 
related to its first-order derivative with respect to some  thermodynamic variable, see e.g. Figure~\ref{OPvar}.
For a finite lattice volume $N_4$ one can only determine a position of the (volume dependent) pseudo-critical point. Positions of such points may in general depend on the order parameter or the method used. Only in $N_4 \to \infty$ limit they must coincide. Let  $\Delta^c(N_4)$ be the pseudo-critical value of the thermodynamic variable $\Delta$, e.g. the coupling constant, measured for a given phase transition for the lattice volume $N_4$. The typical (large)  volume dependence  is 
\beql{scaling}
\Delta^c(N_4) = \Delta^c(\infty) - \frac{C}{N_4^{1/\nu}}\ ,
\eeq
where the critical exponent $\nu$ is  one for a first-order transition and larger than one for a higher-order transition. Thus by making a series of measurements of $\Delta^c(N_4)$ for different lattice volumes $N_4$  one can
establish a value of the critical exponent $\nu$ and in effect determine the order of the phase transition.

Another way of distinguishing between the first- and the higher-order phase transitions in numerical Monte Carlo studies is to analyze the behaviour of the order parameter(s) measured precisely at (or in practice as close as possible to) the transition point.  For a first-order transition the discontinuity of an order parameter  can appear  in its MC history as jumps between two different  states. In such a case, the histogram of the order parameter measured at the pseudo-critical point should show two separate peaks centered around the values generic for the two different phases. Here one  should also  carefully analyze finite size effects related to the finite lattice volume $N_4$ fixed in the numerical studies. The separation of the peaks in the MC history histogram can either increase or decrease with the lattice volume which can imply the first- or the higher-order transition, respectively. 
 If the separation of the  states, generic for the first-order transition, is large enough one typically observes a hysteresis at the  transition region. In order to check that, one can run two separate series of Monte Carlo   simulations,  one  initiated with configurations generic for one phase and the other one initiated with  configurations generic for the other phase. If hysteresis is present then one can observe a (statistically) different behaviour of the two series in the transition region, e.g. the pseudo-critical points measured in the two different series could be shifted versus each other. If hysteresis is absent the results of the two series should (statistically) agree. Running two independent series initiated with different staring configurations is also a good way of checking  {\it thermalization} of the Monte Carlo data, i.e. checking if the MC simulation has run for long enough to reach the proper statistical equilibrium and thus if measurement data can be collected, see e.g. Figure \ref{thermalization}.   

\begin{figure}[H]
\centering
\includegraphics[width=0.53\linewidth]{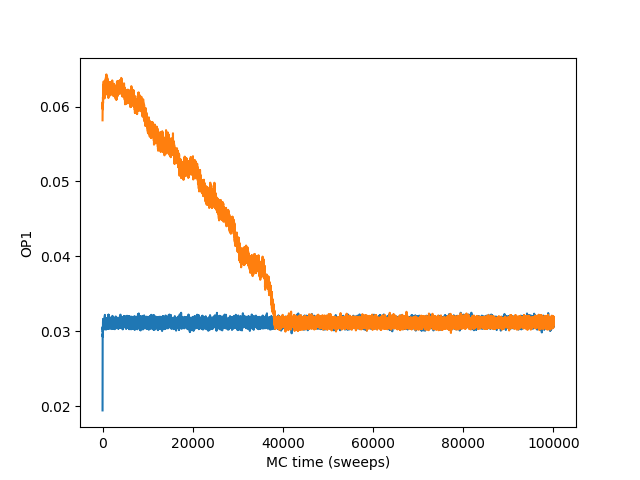}
\caption{\small Thermalization check of  Monte Carlo data series. The plot shows the $OP_1$  order parameter (for definition see equation \rf{OPs}) measured in two independent MC simulations of CDT with toroidal spatial topology with exactly the same parameters, i.e. $N_{4,1}= 300\mathrm{k}$, $T=4$, $\kappa_0=2.2$, $\Delta=0.048$. One simulation was initiated with a configuration from phase $B$ (blue line) and the other one started from a configuration from phase $C_b$ (orange line). Both data series  statistically agree from  ca $ 40000$ sweeps ($1$~sweep = $10^7$ attempted MC moves). Data from earlier MC time history, called the {\it thermalization} period, are excluded from final measurements.}
\label{thermalization}
\end{figure}

Another quantity of interest is the Binder cumulant\footnote{Note that here we use a definition of the Binder cumulant which is shifted (by a  $-2/3$ constant) versus the original Binder's formulation  \cite{Binder:1981a,Binder:1981b,Binder2010}: $B_x=1- \frac{1}{3} \frac{\langle x^{4} \rangle}{\langle x^2\rangle^{2}}$. The definition \rf{binder} was  used in previous CDT phase transition studies \cite{Ambjorn:2011cg,Ambjorn:2012ij,Ambjorn:2019pkp} and thus we keep it in order to ease comparison with these results. The virtue of using our definition is that, as explained in the text, the  deviation of (critical) $B_{OP}$ from zero with rising lattice volume may signal a first order transition, while the convergence to zero is  characteristic of a higher order transition. One could as well use the original Binder's definition and look at the deviation from $2/3$. }
\begin{equation}\label{binder}
B_{OP}\equiv\frac{1}{3}\left(1-\frac{\langle OP^{4} \rangle}{\langle OP^2\rangle^{2}}\right)=-\frac{1}{3}\frac{\langle(OP^2)^2\rangle-\langle OP^2\rangle^{2}}{\langle OP^2\rangle^{2}},
\end{equation}
which is always non-positive because $\langle(OP^2)^2\rangle-\langle OP^2\rangle^{2}\geq 0$,
and it reaches a minimum at the pseudo-critical point $\Delta^c(N_4)$, because there fluctuations are maximal.
In the numerical MC simulations one can measure the (volume dependent) value of the Binder cumulant minimum 
\beql{Bmin}
B^{min}_{OP}(N_4)=B_{OP}(\Delta^c(N_4)) 
\eeq
for different (fixed) lattice sizes $N_4$ and then analyze its behaviour in the large volume limit $N_4\to \infty$.
In the case of a higher-order phase transition the probability distribution of the order parameter $OP$ approaches a Dirac delta around $\langle OP\rangle$ in the infinite volume limit. And then $B^{min}_{OP}(\infty)$ should equal $0$. In the case of the first-order  transition the distribution of the parameter $OP$ is a sum of two distributions centered at  expectation values characteristic for the two different phases. In the infinite volume limit, when these distributions
approach Dirac delta functions, the minimum of the Binder cumulant becomes:
\beql{BminModel}
B^{min}_{OP}(\infty)=-\frac{\langle OP_B\rangle^2+\langle OP_{C_b}\rangle^2}{12 \langle OP_B\rangle^2 \langle OP_{C_b}\rangle^2}
\eeq
where $\langle OP_B\rangle$ and $\langle OP_{C_b}\rangle$ are expectation values of the observable $OP$ at two different phases, say "$B$" and "$C_b$", and the
relative strength of Dirac delta functions is assumed to be  $\frac{\langle OP_B\rangle^2}{ \langle OP_B\rangle^2 + \langle OP_{C_b}\rangle^2}$
 and $\frac{\langle OP_{C_b}\rangle^2}{ \langle OP_B\rangle^2 + \langle OP_{C_b}\rangle^2}$, respectively.

In  Table \ref{TableMethods} we summarize  methods used in  numerical MC simulations of lattice field theories to distinguish between the first- and the higher-order phase transitions. We will then apply these methods in Section \ref{results} to analyze the $B-C_b$ transition in CDT with the toroidal topology of spacial slices.

\begin{table}[H]
\begin{center}
\begin{tabular} {|c|c|c|c|}
\hline
{OBSERVABLE}	& {First-order transition}	& {Higher-order transition} \\ \hline
\hline
Critical exponent $\nu$ in 	&  $\nu$	&  $\nu$ \\
scaling of $\Delta^{c}(N_4)$, eq. \rf{scaling}	&  $=1$	& $>1$  \\ \hline
$OP$ histograms measured at	&  double peaks	& single peak or  \\ 
 pseudo-critical points $\Delta^{c}(N_4)$	&  peak separation $\uparrow$ with $N_4\to\infty$	&  peak separation $\downarrow$  with $N_4\to\infty$  \\ \hline
 Hysteresis of MC data near&  YES & NO   \\ 
  pseudo-critical points $\Delta^{c}(N_4)$	&  hysteresis $\uparrow$ with $N_4\to\infty$	&  or hysteresis $\downarrow$  with $N_4\to\infty$  \\ \hline
Binder cumulant \rf{binder}  	&   $B_{OP}^{min}(N_4\to\infty)$	&   $B_{OP}^{min}(N_4\to\infty)$ \\
minima for $N_4 \to\infty$ &  	$ < 0 $ &  $ =0$ \\ \hline
\end{tabular}
\caption{\small 
Characteristics of the first- and the higher-order phase transitions in MC studies.}\label{TableMethods}
\end{center}
\end{table}

\end{section}

\begin{section}{The properties of the bifurcation phase $C_b$}

The existence of the bifurcation phase in the CDT model with a spherical spatial topology was discovered relatively
late \cite{Ambjorn:2014mra,Ambjorn:2015qja,Ambjorn:2017}. The reason why in the early studies only three phases were discussed was that the basic 
observable used in these approaches was the (average) spatial volume profile of configurations. 
A typical setup for
numerical experiments was to use systems periodic in time, with a period $T$ usually in the range 40 -- 80.
Using the spatial volume 
observable, the three phases, $A$, $B$ and $C$, were characterized by completely different qualitative
behavior. The phase $A$ was characterized by large fluctuations of the 
spatial  volume in the neighboring
time slices. The observed average volume distribution in time corresponded to the unbroken symmetry of the time translations.
In the phase $B$ almost all spatial volume (except for the {\it stalk}, necessary to satisfy the periodic
boundary conditions) was concentrated at a single time slice. This meant that for typical
states in this phase the symmetry of the time translations was fully broken.
The physically most interesting was the phase $C$, where
the volume profile contained the {\it  blob} and the {\it stalk}, again meaning that for a typical configuration
the symmetry of the time translations was  broken. Average volume distribution in the blob and its
fluctuations could be
very accurately explained using the effective mini-superspace model for the isotropic four-dimensional
Euclidean space-time \cite{Ambjorn:2007jv,Ambjorn:2008wc,Ambjorn:2011ph}. Most results were obtained for a particular point in the coupling constant space with
$\kappa_0=2.2$ and $\Delta=0.6$, where it was shown that volume distribution scaled with the
total $N_{4,1}$ lattice volume in a way consistent with the Hausdorff dimension $d_H=4$.

Similar measurements performed for decreasing values of $\Delta$ showed that, although qualitatively the
volume profile still contained a blob and the stalk, the scaling properties did not follow those
determined in the de Sitter phase $C$. It was observed that the scaling was consistent with that predicted for
systems with the Hausdorff dimension $d_H=\infty$. The name {\it bifurcation phase} $C_b$ appeared to describe the
additional property observed in the volume profile: a different behavior in the even and odd time slices when the  time period $T$ was sufficiently small 
\cite{Ambjorn:2014mra}.
It was soon realized that the reason for the observed behavior came from the breaking of the isotropy of the
spatial volume distribution in the new phase. For the time slices separated by two units in time, vertices
with very high coordination numbers appeared, leading to a formation of highly nontrivial geometric objects, forming
a chain in the time direction. A physical interpretation of these objects was conjectured to be a result
of a local signature change from Euclidean to Lorentzian \cite{Ambjorn:2015qja}, producing objects with some qualitative
similarity to a {\it black hole} or rather a series of {\it black points}. A detailed description
of the microscopic mechanism producing such effects will be the subject of a separate paper. 

As can be seen in Figure \ref{pdnew}, for decreasing values of $\Delta$ and a fixed value of $\kappa_0$,
one observes a phase transition between the $C_b$ and $B$ phases. The properties of this phase transition
were very accurately measured in the case of a spherical spatial topology \cite{Ambjorn:2011cg,Ambjorn:2012ij,Ambjorn:2017},
although originally the phase $C_b$ was interpreted as being a part of the {\it de Sitter}
phase $C$. Results indicated that the phase transition was higher order, a very important
property from a theoretical point of view, as explained earlier. The purpose of the present analysis
is to check if the position and properties of the phase transition remain the same for systems with
the spatial topology $\Sigma$ of a sphere $S^3$ and of a three-torus $T^3$.

The first question to be asked is: are the qualitative properties in the $C_b$ phase similar or different
when we consider systems with a different spatial topology. Again we may look at the simplest object, a volume
profile for systems with the periodicity $T$ of the same order as the one used in the spherical case.
This is the observable which was found to behave differently in the $C$ phase. The observed volume profile, in this
case, was found to be flat rather than containing a blob \cite{Ambjorn:2016fbd,Ambjorn:2017226}. The reason of such a behavior 
could be explained using a mini-superspace spatially isotropic model for a system with the spatial topology of
a three-torus. The averaged volume profile is flat since in the toroidal case the time translation symmetry
remains unbroken \cite{Ambjorn:2016fbd,Ambjorn:2017226}.

Investigations show that this is not the case in the bifurcation phase $C_b$. The volume profile observed for the
point in the coupling constant space, typical for the bifurcation phase ($\kappa_0=2.0$ and $\Delta=0.2$)
shows the appearance of a blob and the stalk, see Figure \ref{figbif}, the same way as it was observed in the spherical case. %
Also the scaling of the volume profile with the total $N_{4,1}$ lattice volume is consistent with the Hausdorff dimension $d_H=\infty$, the same as in the spherical CDT.
The analysis of the geometric properties of configurations in the bifurcation phase $C_b$ shows that also
from a microscopic point of view the toroidal and spherical cases are very similar. In both topologies,
we observe the high-order vertices, separated in time by two steps. The shape of the blob observed for 
periodicity $T$ large enough ($T\geq 20$) again scales consistently with the infinite Hausdorff
dimension. The difference is observed in the stalk, which has a much larger volume for a torus than that for a
sphere. This is well understood and results from the fact that a minimal $3D$ spatial configuration
depends strongly on the topology (see \cite{Ambjorn:2016fbd}).

As a conclusion, one may expect the critical properties of the phase transition between the $C_b$ and $B$
phases to be very similar in both topologically different realizations of the model. Below
we show that this is indeed the case. The measurement of the critical behavior on the boundary 
between $C_b$ and $C$ phases may, on the other hand, be different, or at least difficult to be determined
numerically.

\begin{figure}[H]
\centering
\includegraphics[width=0.4\linewidth]{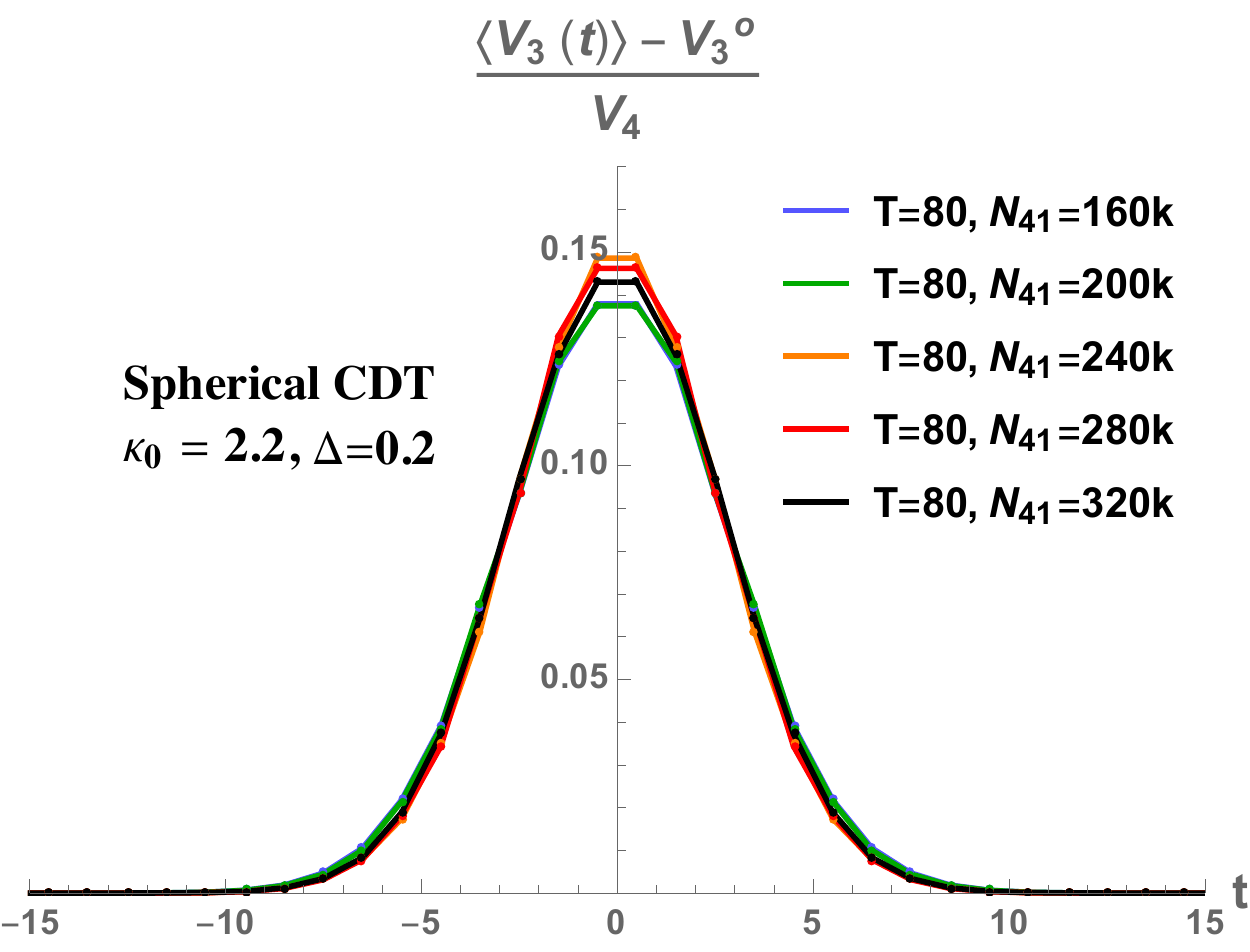}
\includegraphics[width=0.4\linewidth]{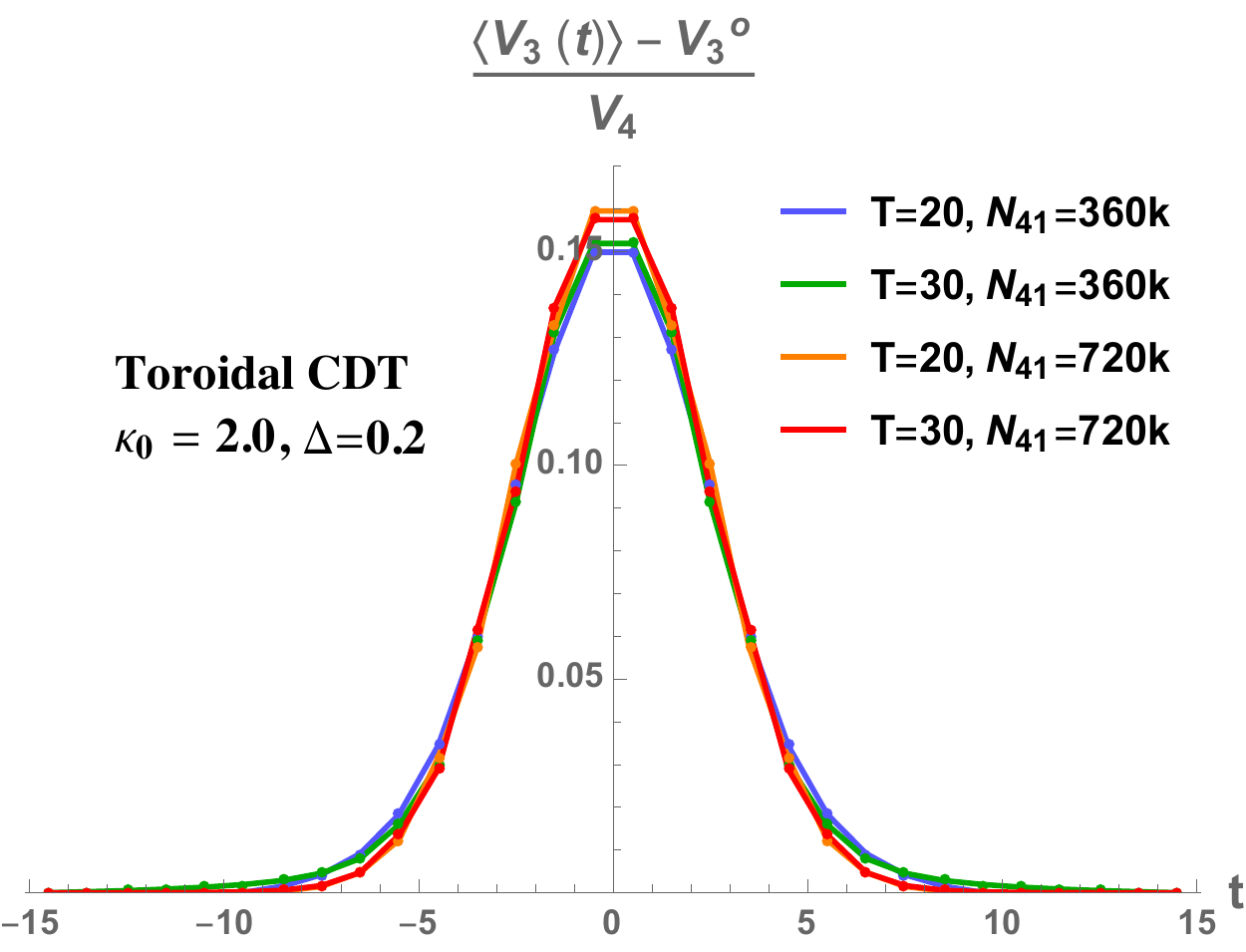}
\caption{\small The (rescaled) average spatial volume profiles $\langle V_3(t)\rangle $ observed in the bifurcation phase $C_b$ in the spherical (left plot) and the toroidal (right plot) CDT. In both plots the spatial volume profiles were presented with respect to the centre of volume, set at $t=0$, and shifted  by a (constant  $V_3^0$) volume measured in the {\it stalk} range ($|t|>\sim 10$),  $V_3^0$ being different for each volume profile (in general  $V_3^0$ is bigger in the toroidal CDT where discretization effects are larger). Data measured for various total $N_{4,1}$ lattice volumes and different $T$ were rescaled by $V_4=\sum_t(\langle V_3(t)\rangle - V_3^0$), i.e. in agreement with the Hausdorff dimension $d_H=\infty$.}
\label{figbif}
\end{figure}

\end{section}

\begin{section}{The $B-C_b$ phase transition in the toroidal CDT}\label{results}

Below, we present the results of the $B-C_b$ phase transition study in CDT with the toroidal spatial topology. The $B-C_b$ transition was earlier studied in the spherical spatial topology \cite{Ambjorn:2011cg,Ambjorn:2012ij,Ambjorn:2017} where it was classified to be the higher order transition. As explained in Section \ref{method} in order
to investigate the phase transition one has to make a series of Monte Carlo simulations for various points in the CDT $(\kappa_0,\Delta)$ parameter space\footnote{In each Monte Carlo simulations the $\kappa_4$ is fine-tuned to the critical value, which depends on  $\kappa_0$ and $\Delta$ and also on the lattice volume $N_{4,1}$.}, around the phase transition point. In this study all measurements were taken for one fixed value of $\kappa_0=2.2$ and  for a sequence of $\Delta$ values.\footnote{The same $\kappa_0$ value was earlier used in the $B-C_b$ transition studies in the spherical CDT.} In each simulation the $N_{4,1}$ lattice volume  of the system (i.e. the total number of $\{4,1\}$ and $\{1,4\}$ simplices) is fixed or, more precisely, it fluctuates around the target value $\bar N_{4,1}$. The lattice volume is controlled by a volume-fixing potential 

\beql{volfix}
\delta V = \epsilon (N_{4,1} -\bar N_{4,1})^2
\eeq

\noindent added to the bare Einstein-Hilbert-Regge action of CDT \rf{bareaction}
such that the volume is sharply peaked around a chosen value of $\bar N_{4,1}$, with a well-defined amplitude of fluctuations    {$\propto 1/ \epsilon$}. In  the CDT  Monte Carlo simulations one also has to set the length of the (periodic) time axis, i.e. the number of (integer) time slices $T$. In our case the number of time slices was equal $T= 4${, the numerical constant governing the magnitude of volume fluctuations was fixed at  $\epsilon = 0.00002$ and measurements were performed every $10^7$ attempted Monte Carlo moves (such that the measured   $N_{4,1}$ volume could differ from the target $\bar N_{4,1}$ volume)}.\footnote{In principle  MC simulation results could depend on the  set of parameters used, such as the volume fixing method (one could e.g. fix the total $N_4$ volume instead of the $N_{4,1}$ volume) or the number of time slices $T$ but as advocated in \cite{Ambjorn:2019pkp} the order of CDT phase transitions does not depend on that.}

In our analysis we will focus on the behaviour of four order parameters which have previously been successfully used in  phase transition studies both in the spherical \cite{Ambjorn:2012ij,Coumbe:2015oaa,Ambjorn:2017tnl} and the toroidal \cite{Ambjorn:2018qbf,Ambjorn:2019pkp,Ambjorn:2019lrm} CDT,\footnote{Here we use a slightly different definition of $OP_1$ than in previous CDT phase transition studies, where it was: $OP_1\equiv N_0 / N_4$. Current definition is more natural when $N_{4,1}$ volume is fixed (see equation \rf{volfix}) which was the case in all MC simulations described herein.}

\beql{OPs}
\begin{array}{l}
OP_1={N_0}/ { N_ {4,1}}, \hspace{3.5cm} OP_2={N_ {3,2}}/{N_ {4,1}}, \\
\\
OP_3=\sum_t(V_3(t+1)-V_3(t))^2,  \quad  \quad OP_4= \max_v O(v),
\end{array}
\eeq 
where $V_3(t)$ is the spatial volume\footnote{To ensure consistency with our earlier publications we define $V_3(t)$ as \underline{twice} the number of spatial tetrahedra with the integer time coordinate $t$.} in the time slice $t$ and $O(v)$ is the vertex coordination number, i.e. the number of simplices sharing a given vertex $v$. 
The behaviour of the order parameters in all CDT phases has been summarized in Table \ref{tab1}. Specifically when changing from the phase $B$ to the phase $C_b$ the $OP_1$, $OP_2$ and $OP_4$ increase in value while the $OP_3$ decreases, see  Figure \ref{OPmean}.
\begin {table}[H]
\begin{center} 
\begin{tabular} {|c||c|c|c|c|}
\hline
  & Phase  $ {A}$& \ Phase  $ {B}	$ &Phase  $ {C}$ & Phase  $ {C_{b}}$\\ \hline
\hline
$OP_1$&large&	small &	medium&	medium 	\\ \hline
$OP_2$&small&	small &	large& 	large	 	\\ \hline
$OP_3$&medium&	large &	small&	medium 	\\ \hline
$OP_4$&small&	large &	small& 	large 	\\ \hline
\end{tabular}
\caption {\small Order parameters used in CDT phase transition studies.} \label{tab1} 
\end{center}
\end{table}
The MC simulations were performed for nine different  (fixed) lattice volumes, i.e. for $\bar N_{4,1}=40\mathrm{k}, 60\mathrm{k}, 80\mathrm{k}, 100\mathrm{k}, 120\mathrm{k}, 140\mathrm{k}, 160\mathrm{k}, 300\mathrm{k}, 400\mathrm{k}$. For each lattice volume $\bar N_{4,1}$ the  approximate location of the $B-C_b$ phase transition point was found and then  a series of precise measurement was performed for $\Delta$ in the range around the expected critical value $\Delta^c$ with a resolution of $0.001$. Each  measurement series was performed twice, each time for a different initial triangulation: one from phase $B$ and   one from phase $C_b$, and the two data series were compared in order to check thermalization  and possible hysteresis, see e.g. Figure \ref{thermalization}.   For each lattice volume $\bar N_{4,1}$ and each of the two measurement series ($s=B,C_b$) and each of the four order parameters $OP_i$ ($i=1,2,3,4$) the precise position of the (volume dependent) pseudo-critical point $\Delta_{i,s}^c(N_{4,1})$ was established based on the peak of the $OP_{i,s}$ susceptibility $\chi_{OP_{i,s}}$, see Figure \ref{OPvar} where we present the results of measurements for the lattice volume $\bar N_{4,1}=100\mathrm{k}$. The values of $\Delta_{i,s}^c(N_{4,1})$   measured for different $OP_i$ and in the two data series in general coincide up to the  used $\Delta$ resolution. If the results for various $OP_i$ or for various data series  are different, usually shifted not more than by the $\Delta$ difference of $0.001$, we simply take the arithmetic mean 
\beql{delcrit}
\Delta^c(N_{4,1})=\frac{1}{8}\sum_{s\in\{B,C_b\}}\sum_{i=1}^4 \Delta_{i,s}^c(N_{4,1})
\eeq   
and assign a correspondingly larger  measurement error, e.g. for the  lattice volume $\bar N_{4,1}=100\mathrm{k}$ one has $\Delta^c(N_{4,1}=100\mathrm{k})=0.0376\pm0.0016$.

\begin{figure}[H]
\centering
\includegraphics[width=0.37\linewidth]{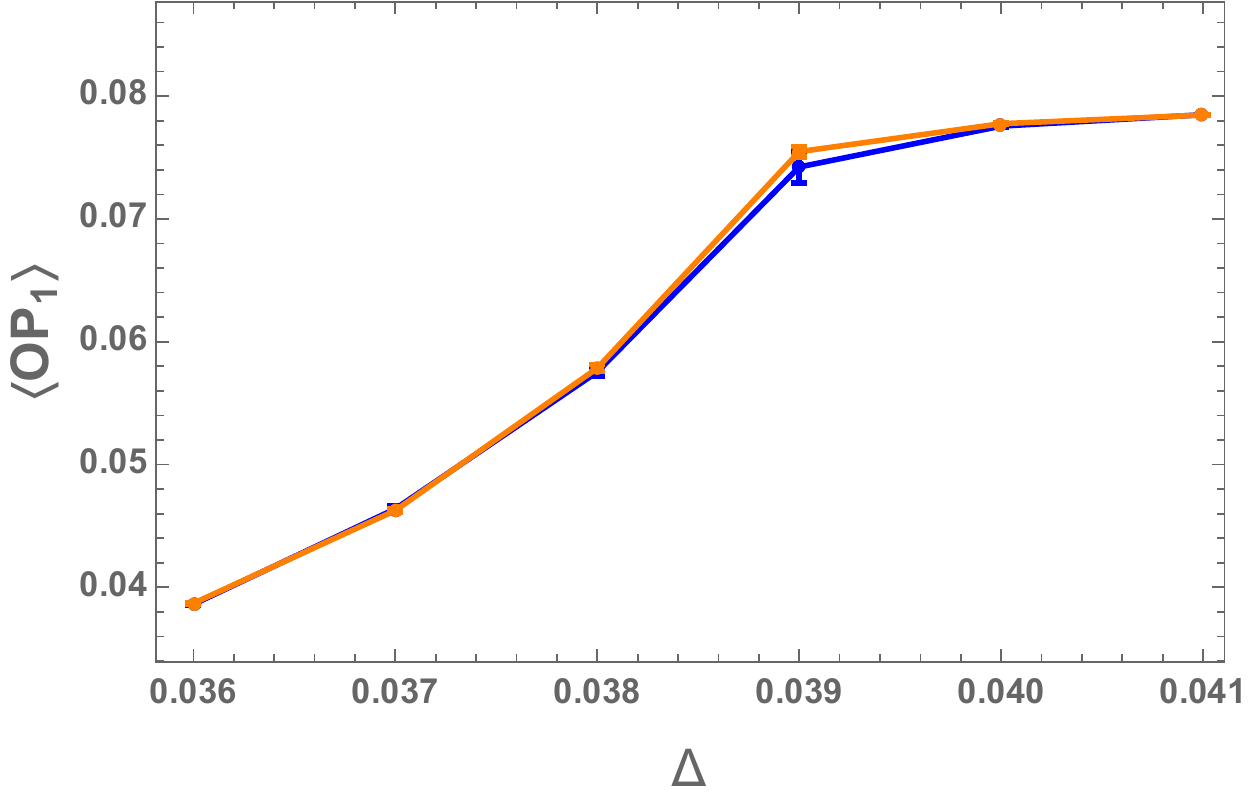}
\includegraphics[width=0.37\linewidth]{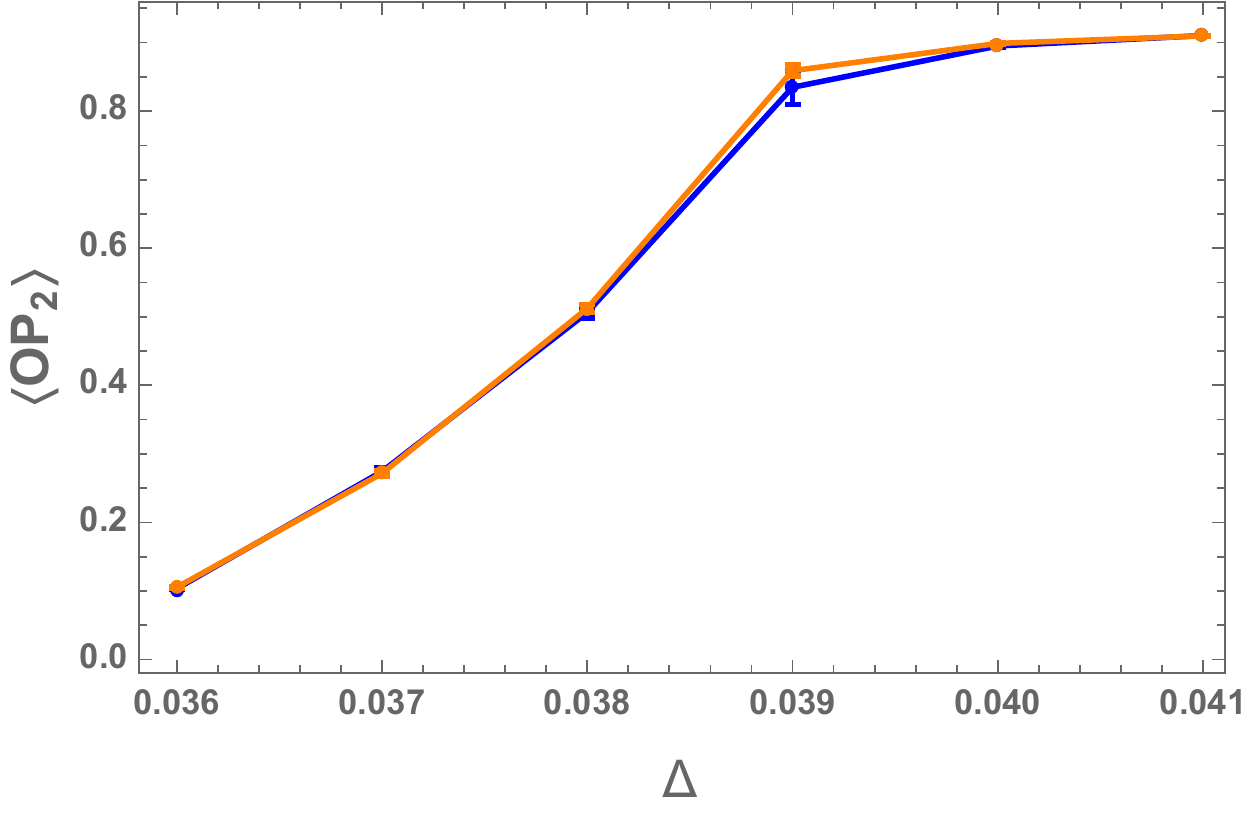}
\includegraphics[width=0.37\linewidth]{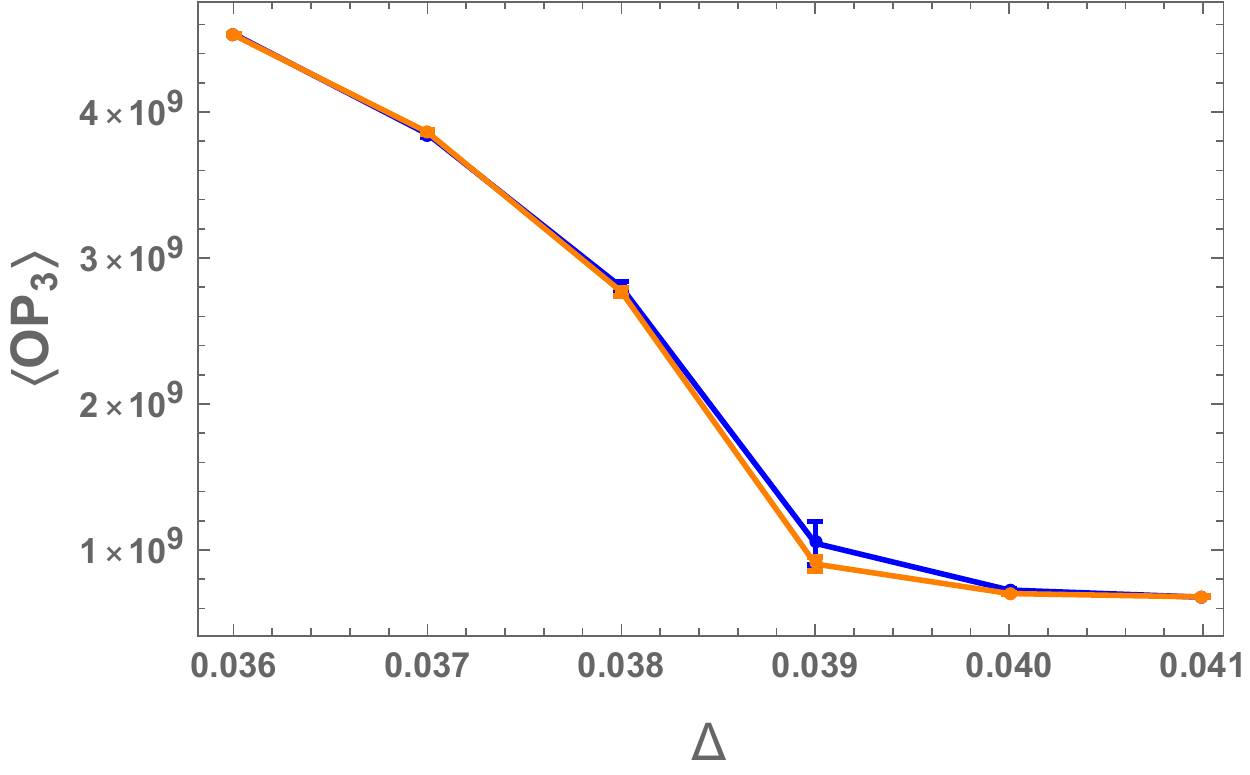}
\includegraphics[width=0.37\linewidth]{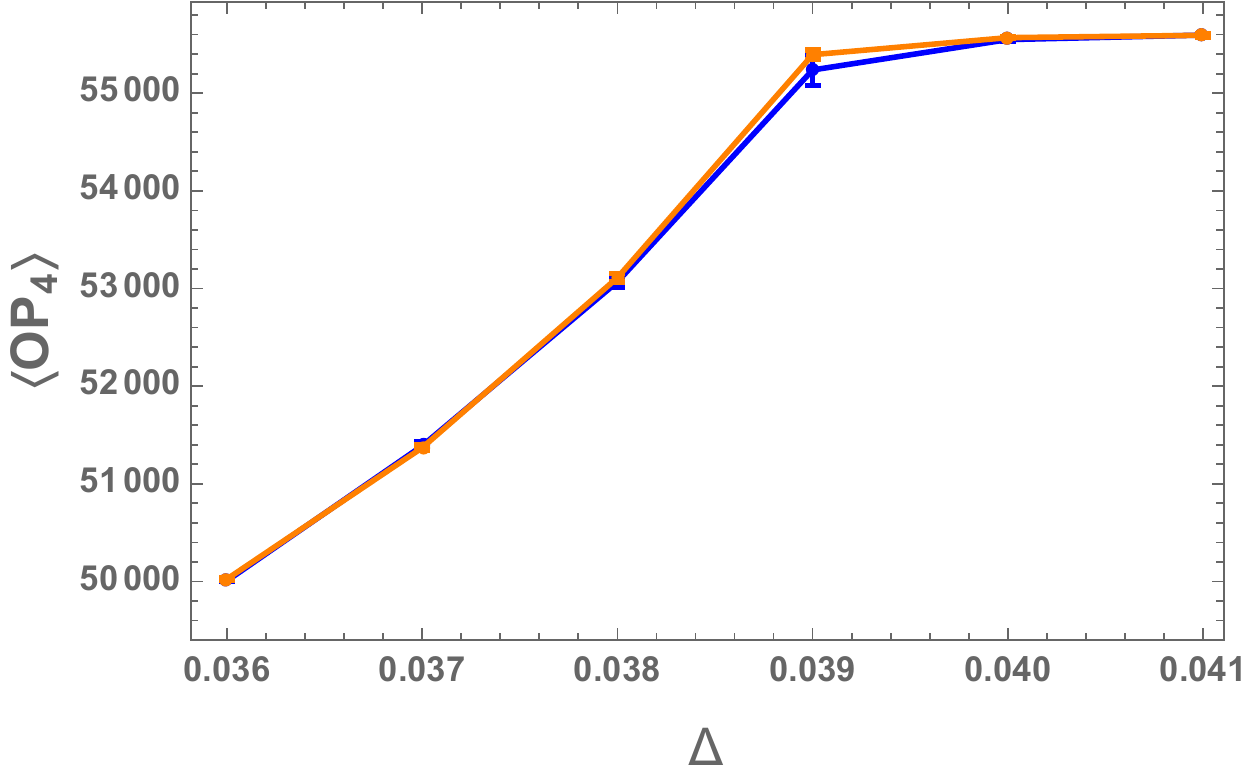}
\caption{\small Mean values of the four order parameters  \rf{OPs} $\langle OP_1\rangle$, ..., $\langle OP_4\rangle$ as a function of $\Delta$ in the $B-C_b$ phase transition region in CDT with toroidal spatial topology for fixed $\kappa_0=2.2$ and the lattice volume $N_{4,1}=100\mathrm{k}$. Blue data points are for the MC series started from a triangulation in phase $B$ while orange data points were started from a triangulation in phase $C_b$. {Error bars were estimated using a single-elimination (binned) jackknife procedure, where the bin sizes were selected in such a way that the statistical errors are maximized.}}
\label{OPmean}
\end{figure}

\begin{figure}[H]
\centering
\includegraphics[width=0.37\linewidth]{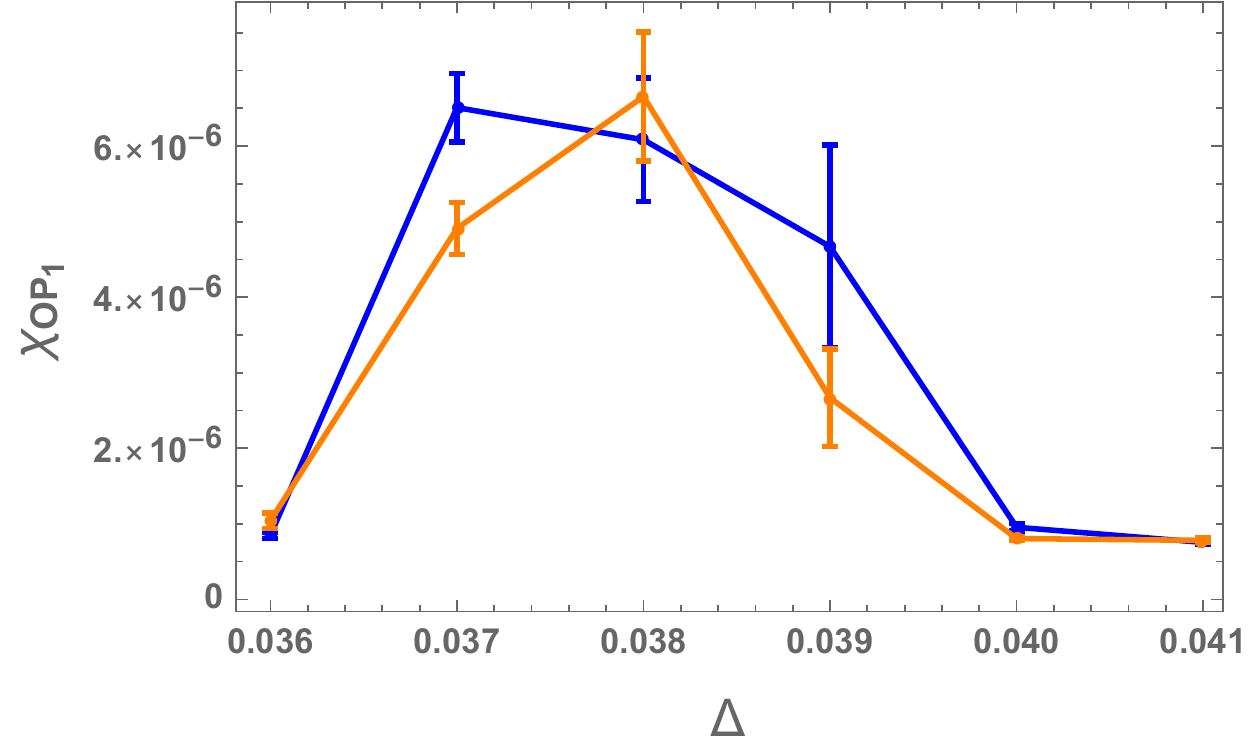}
\includegraphics[width=0.37\linewidth]{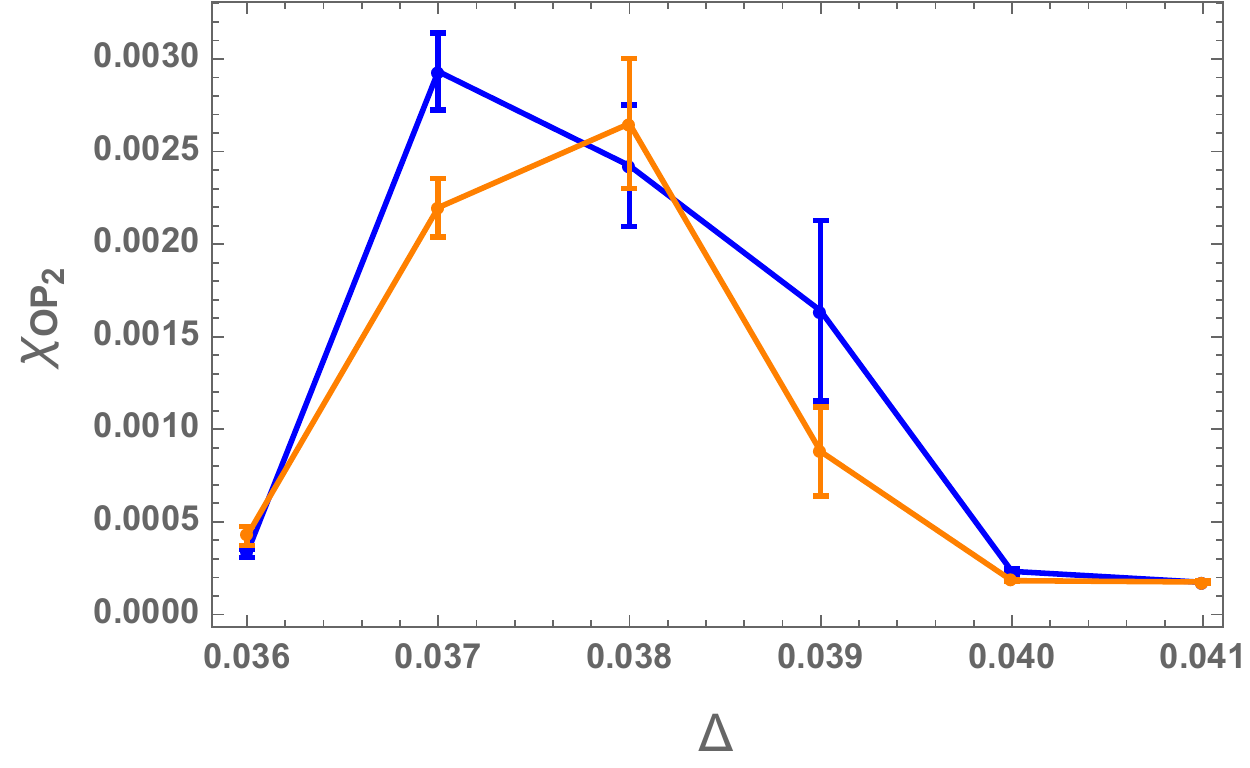}
\includegraphics[width=0.37\linewidth]{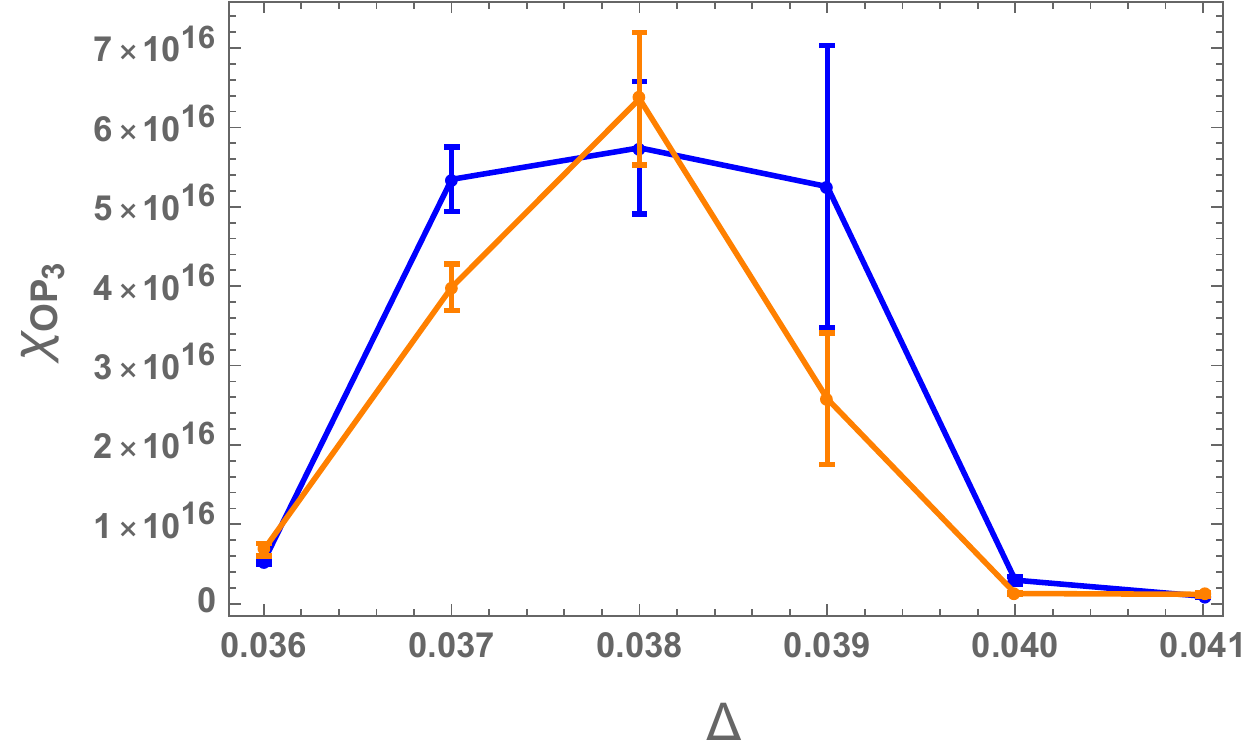}
\includegraphics[width=0.37\linewidth]{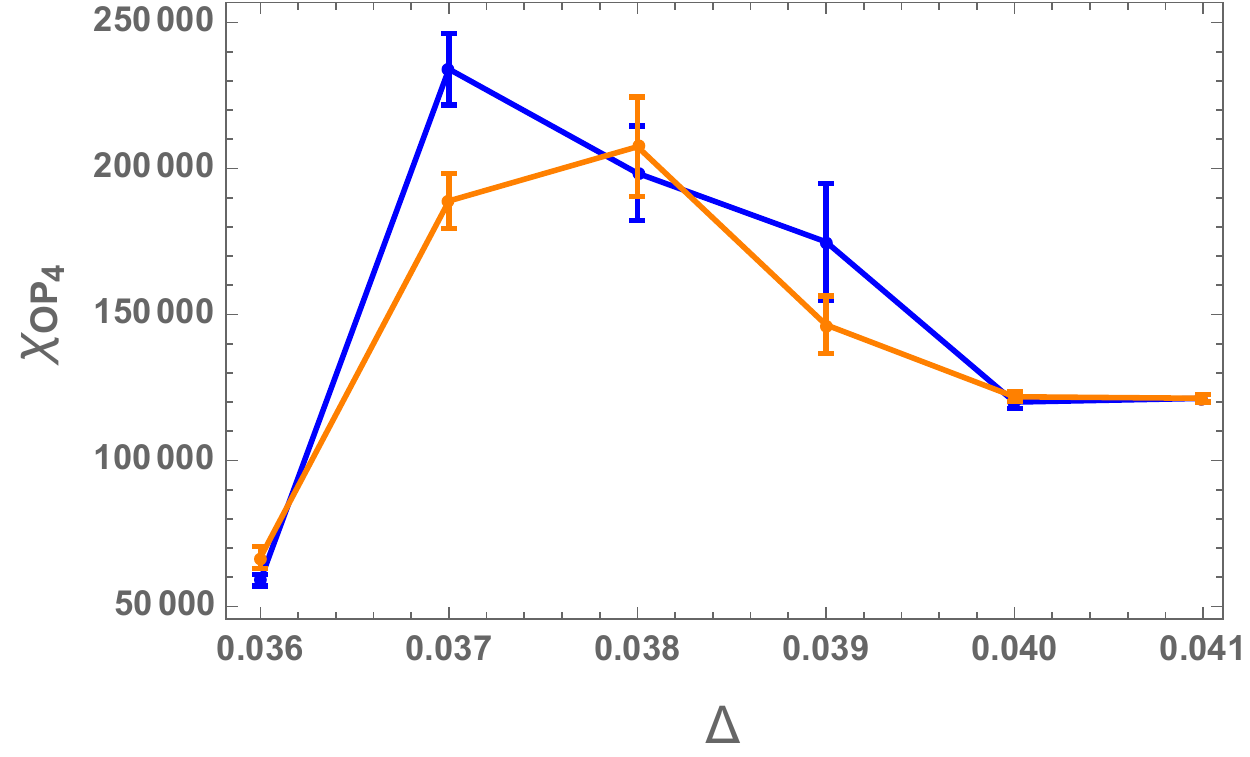}
\caption{\small Susceptibilities \rf{susc} of the four order parameters \rf{OPs} $\chi_{OP_1}$, ..., $\chi_{OP_4}$ as a function of $\Delta$ in the $B-C_b$ phase transition region in CDT with toroidal spatial topology for fixed $\kappa_0=2.2$ and the lattice volume $N_{4,1}=100\mathrm{k}$. Blue data points are for the MC series started from a triangulation in phase $B$ while orange data points were started from a triangulation in phase $C_b$. {Error bars were estimated using a single-elimination (binned) jackknife procedure, where the bin sizes were selected in such a way that the statistical errors are maximized.} }
\label{OPvar}
\end{figure}

Then we fit the finite size scaling relation \rf{scaling} to the measured $\Delta^c(N_{4,1})$ values. The best fit of the true (infinite volume) critical point is $\Delta^c(\infty)=0.073\pm0.004$, and the  best fit  of the critical scaling exponent is $\nu=2.7\pm0.4$ which supports the higher-order nature of the $B-C_b$ phase transition, see also Figure \ref{Figfit} where we plot the measured data together with the best fit of the scaling relation  \rf{scaling} and  compare it to the fit with a forced value of $\nu=1$ (typical for a first-order transition) showing that the quality of the latter fit is much worse. The measured values of the true critical point and the critical exponent 
also agree with $\Delta^c(\infty)=0.077\pm0.004$ and  $\nu=2.51\pm0.03$ measured in CDT with the spherical spatial topology \cite{Ambjorn:2012ij}, giving strong evidence that the results are  independent of the topology chosen (at least for the toroidal and the spherical one).

\begin{figure}[H]
\centering
\includegraphics[width=0.53\linewidth]{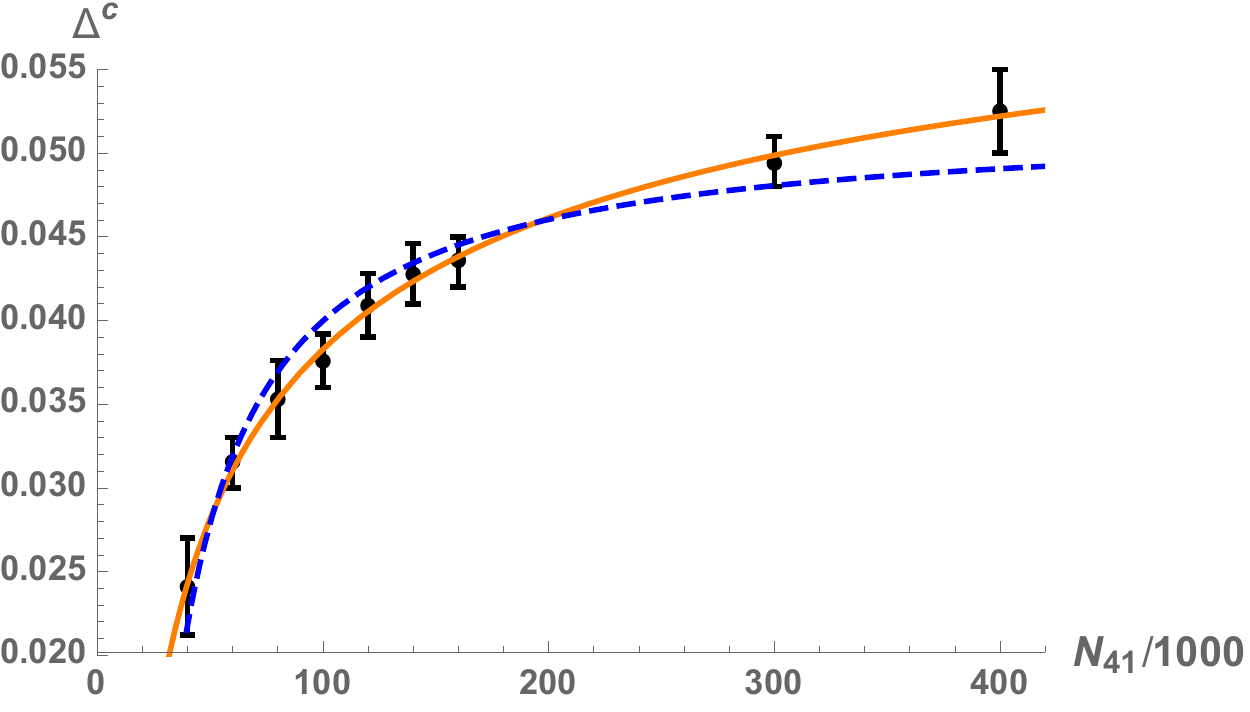}
\caption{\small Lattice volume dependence of the pseudo-critical $\Delta^c(N_{4,1})$ values in CDT with toroidal spatial topology and for fixed $\kappa_0=2.2$ together with the fit of the finite size scaling relation \rf{scaling} with critical exponent $\nu=2.7$ (orange solid line) and the same fit with a forced value of  $\nu=1$ (blue dashed line). }
\label{Figfit}
\end{figure}

\begin{figure}[H]
\centering
\includegraphics[width=0.45\linewidth]{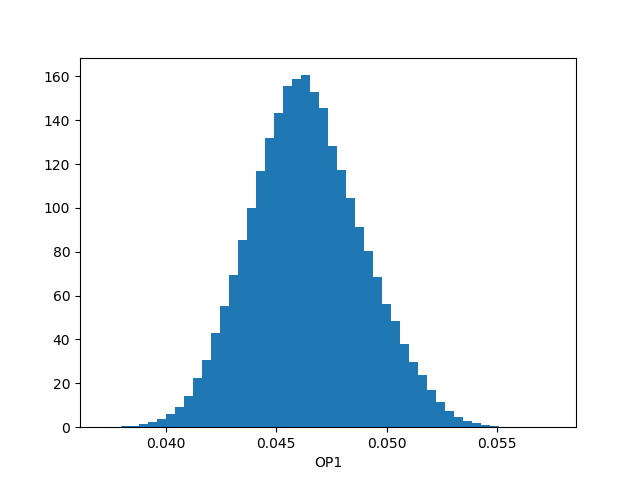}
\includegraphics[width=0.45\linewidth]{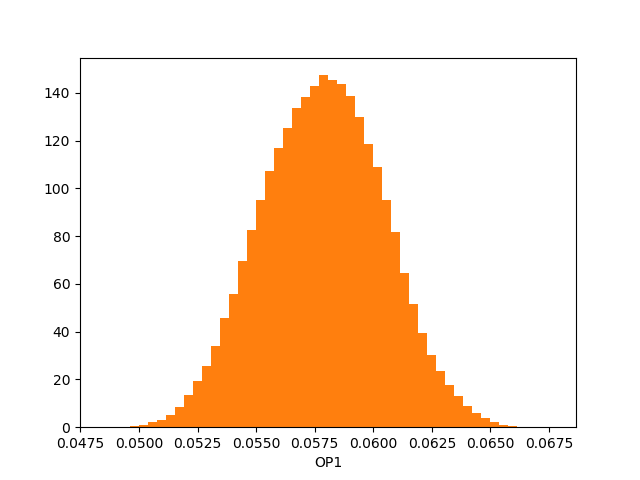}
\caption{\small Histograms of the MC history of the $OP_1$  order parameter \rf{OPs} measured in CDT with toroidal spatial topology for fixed $\kappa_0=2.2$ and the lattice volume $N_{4,1}=100\mathrm{k}$. The left plot is for data series started from configuration in phase $B$ and $\Delta=0.037$ (i.e. the peak of susceptibility $\chi_{OP_{1}}$ measured for this data series, see Figure \ref{OPvar}) while the right plot is for data series initiated in phase $C_b$ and $\Delta=0.038$ (peak of $\chi_{OP_{1}}$ for this data series).}
\label{FigHist}
\end{figure}
\newpage

In order to corroborate  this result, we have performed the detailed Monte Carlo history analysis of all order parameters  at (and in the vicinity) of the measured pseudo-critical points, see Figure~\ref{FigHist} where we plot the MC history histograms of the $OP_1$ measured for the example $N_{4,1}=100\mathrm{k}$  volume and for $\Delta= 0.037$ (peak of $\chi_{OP_{1,B}}$) and $\Delta= 0.038$ (peak of $\chi_{OP_{1,C_b}}$). In none of the cases have we observed the double peaks in the measured histograms nor the hysteresis of the measured data series. These results support the higher-order $B-C_b$ transition.

Finally, we have analyzed the behaviour of the Binder cumulants \rf{binder} in search of minima, see Figure \ref{OPbin} where we plot  data measured for  $N_{4,1}=100\mathrm{k}$. The value of pseudo-critical $\tilde \Delta_{i,s}^{c}(N_{4,1})$ defined by the minimum of the Binder cumulants $B_{OP_{i,s}}$ in general coincides with the $\Delta_{i,s}^c(N_{4,1})$ value defined by the maximum of susceptibility  $\chi_{OP_{i,s}}$, the possible shift is usually up to $\Delta$ difference of $0.001$.  In Figure \ref{N41bin} we plot the measured values of $B^{min}_{OP_{i,s}}(N_{4,1})\equiv B_{OP_{i,s}}(\tilde \Delta_{i,s}^{c}(N_{4,1}))$ as the function of the lattice volume $N_{4,1}$.\footnote{In the plot we skip data measured for $N_{4,1}=400\mathrm{k}$ which can  be not accurate enough as these systems did not thermalize completely resulting in large measurements errors.} All Binder cumulants measured for $OP_1$,..., $OP_4$ visibly grow towards zero when $N_{4,1}$ is increased, which again favours the higher-order nature of the $B-C_b$ transition. 

\begin{figure}[H]
\centering
\includegraphics[width=0.37\linewidth]{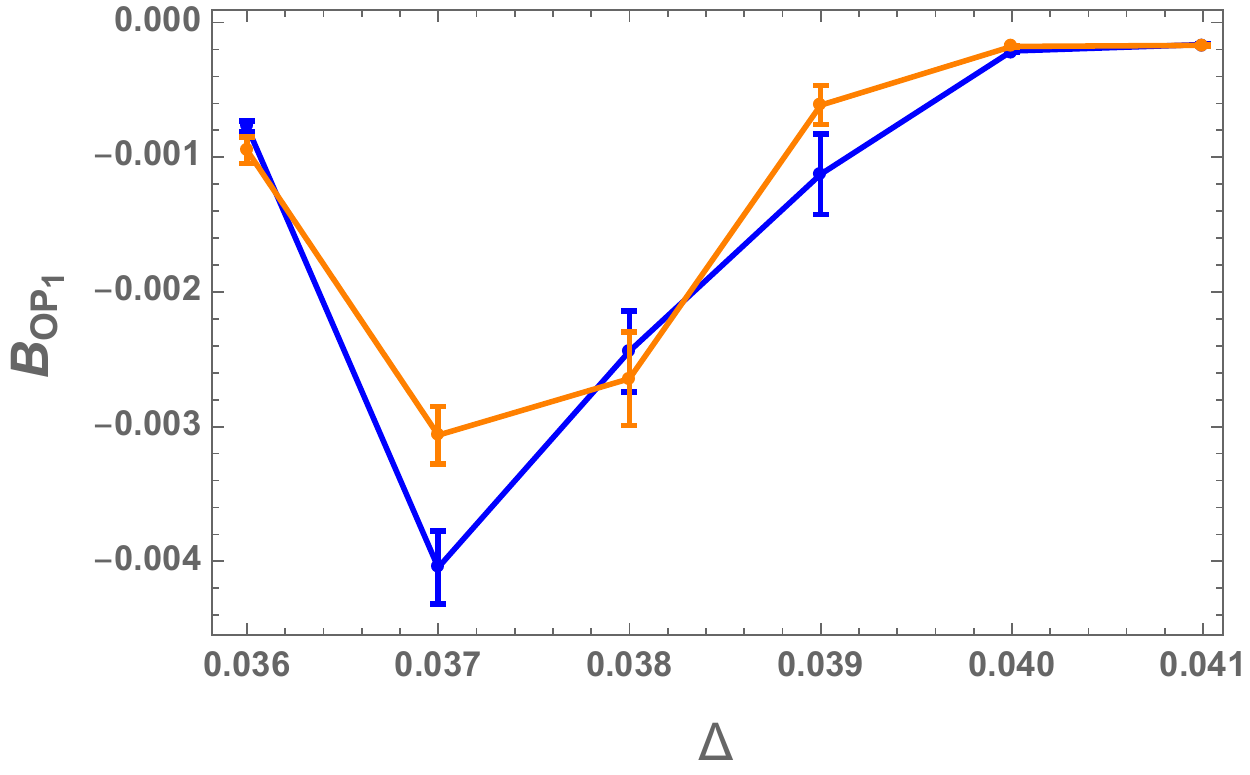}
\includegraphics[width=0.37\linewidth]{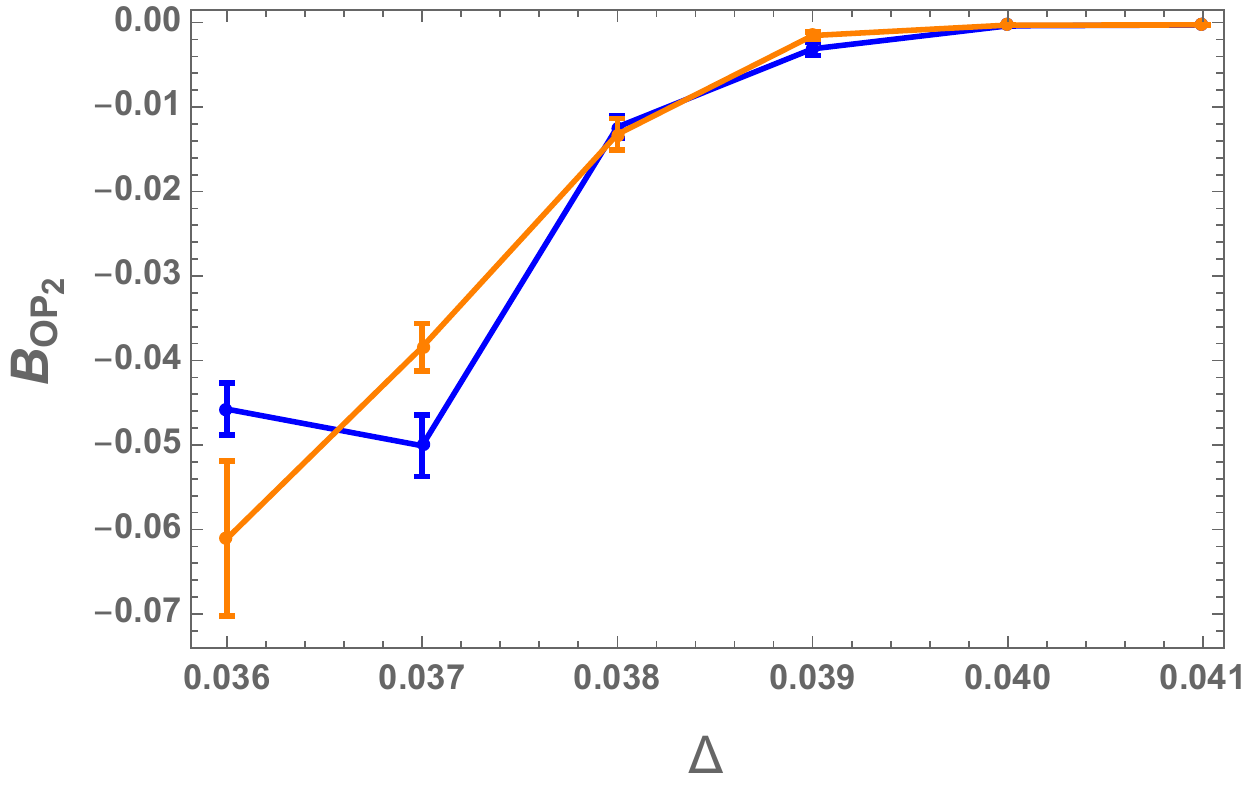}
\includegraphics[width=0.37\linewidth]{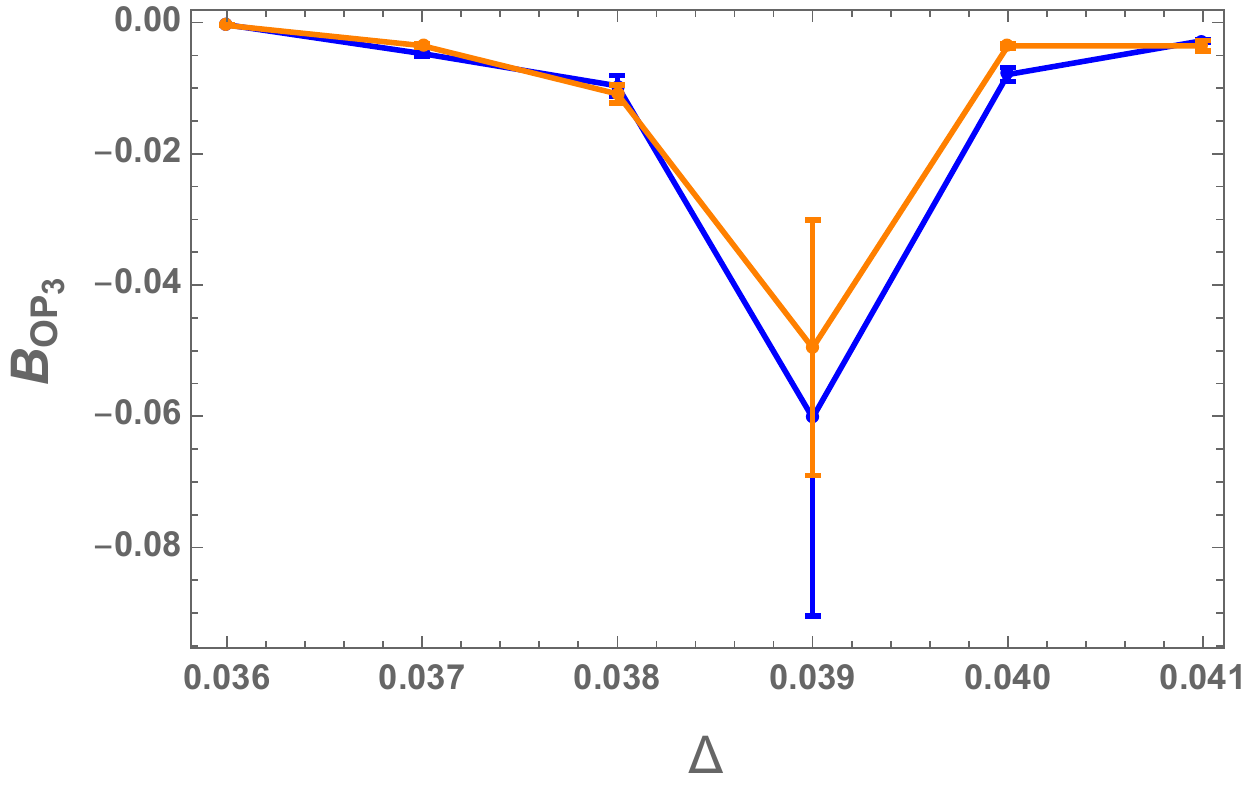}
\includegraphics[width=0.37\linewidth]{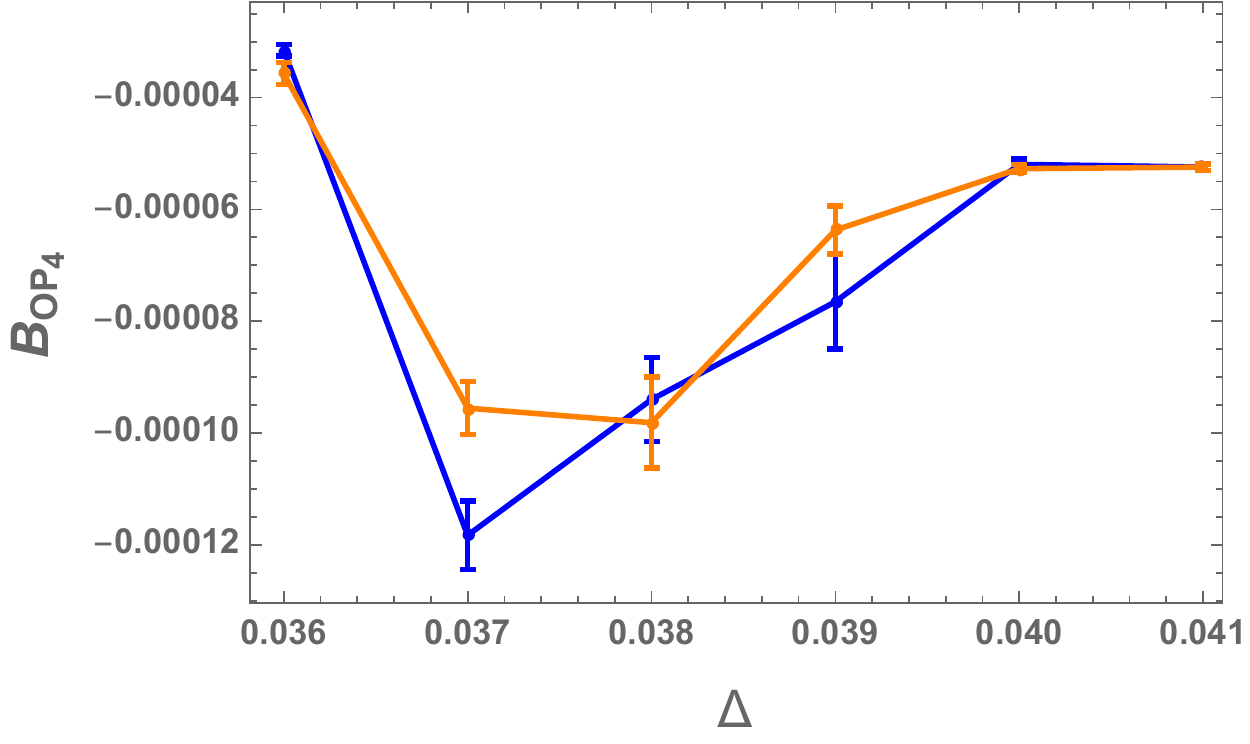}
\caption{\small Binder cumulants \rf{binder} of the four order parameters \rf{OPs} $B_{OP_1}$, ..., $B_{OP_4}$ as a function of $\Delta$ in the $B-C_b$ phase transition region in CDT with toroidal spatial topology for fixed $\kappa_0=2.2$ and the lattice volume $N_{4,1}=100\mathrm{k}$. Blue data points are for the MC series started from a triangulation in phase $B$ while orange data points were started from a triangulation in phase $C_b$. {Error bars were estimated using a single-elimination (binned) jackknife procedure, where the bin sizes were selected in such a way that the statistical errors are maximized.}}
\label{OPbin}
\end{figure}

\begin{figure}[H]
\centering
\includegraphics[width=0.37\linewidth]{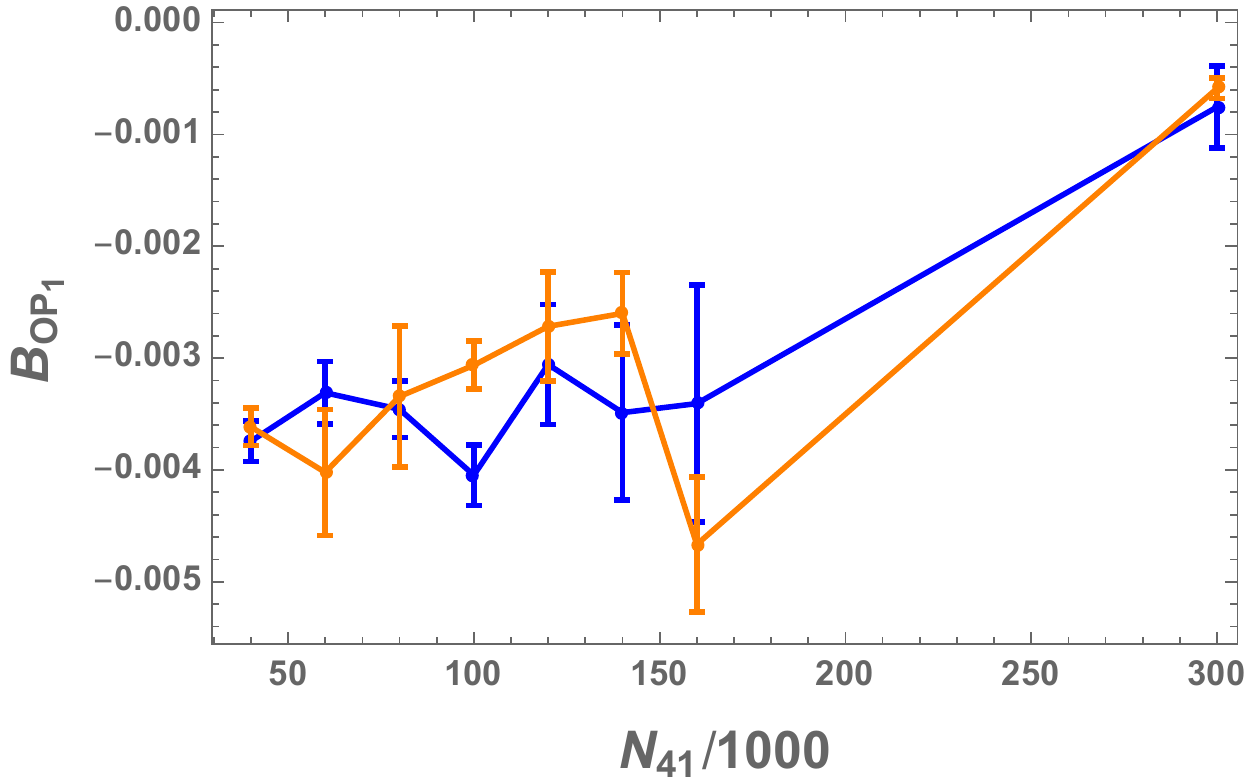}
\includegraphics[width=0.37\linewidth]{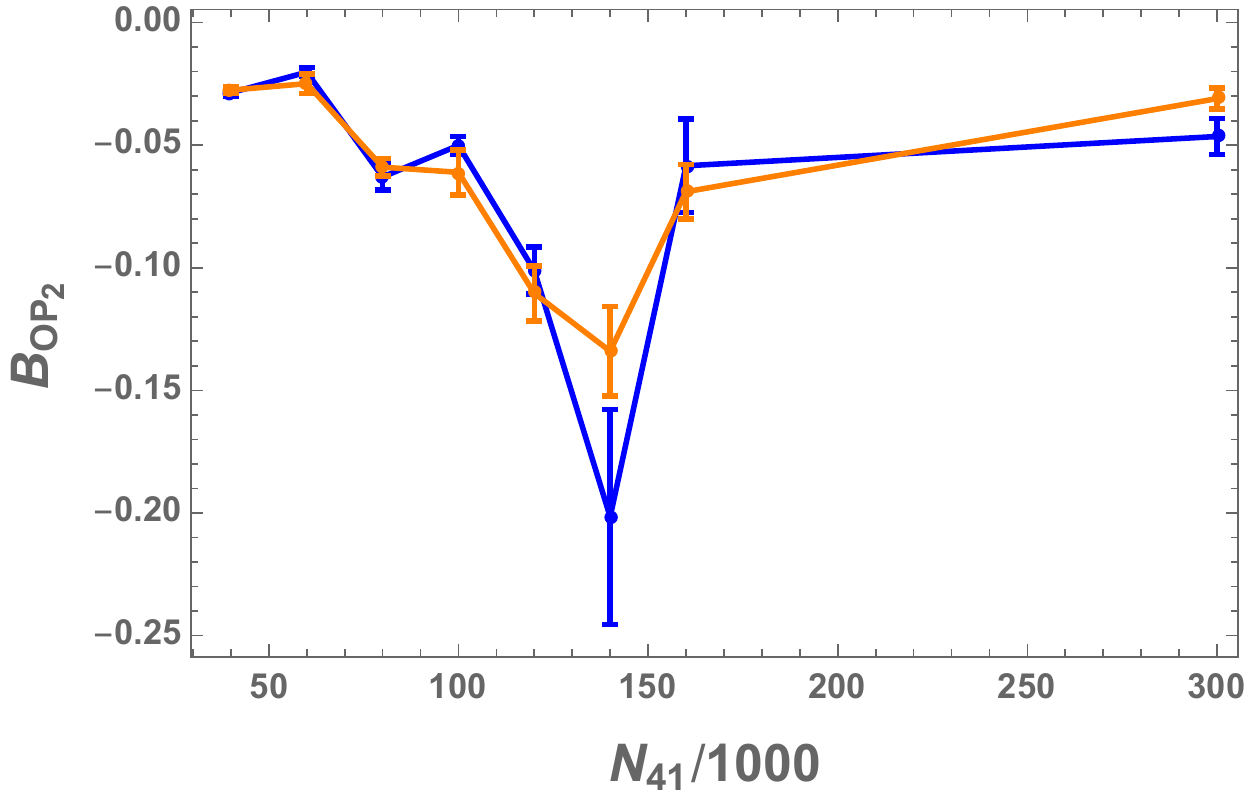}
\includegraphics[width=0.37\linewidth]{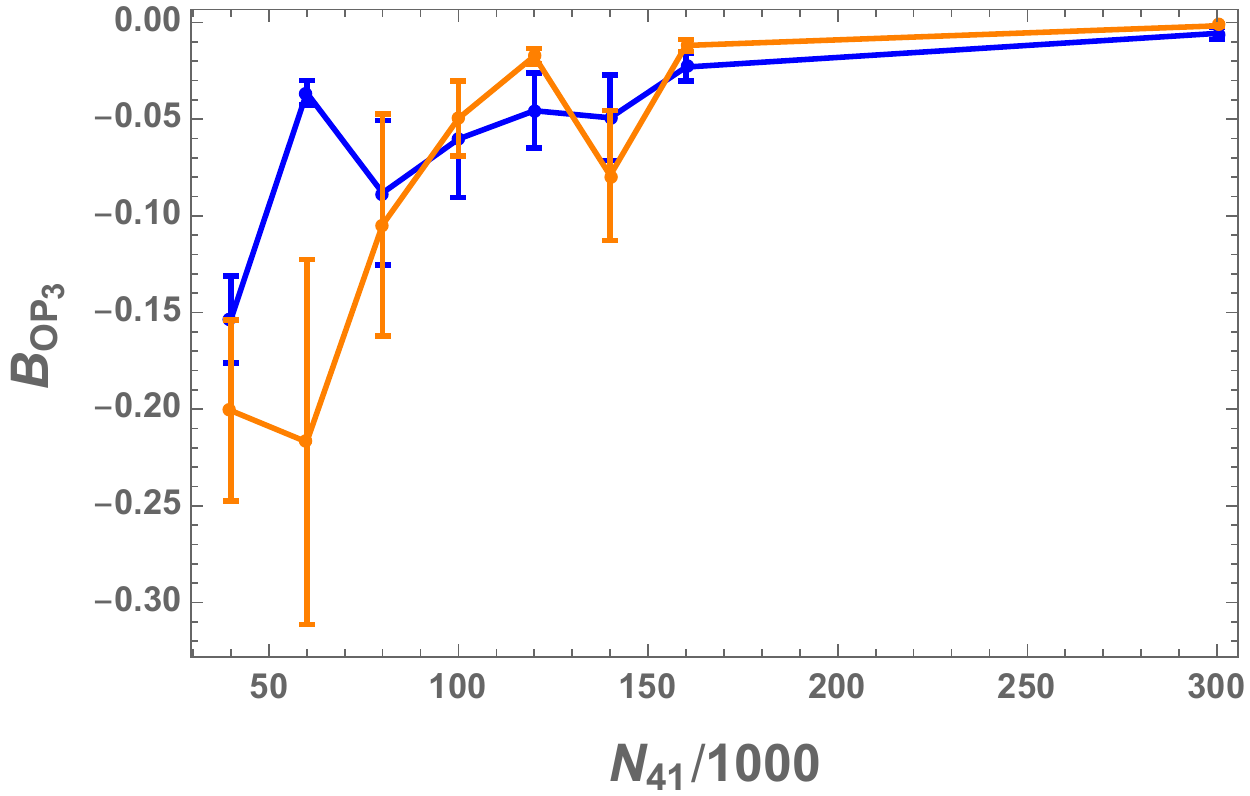}
\includegraphics[width=0.37\linewidth]{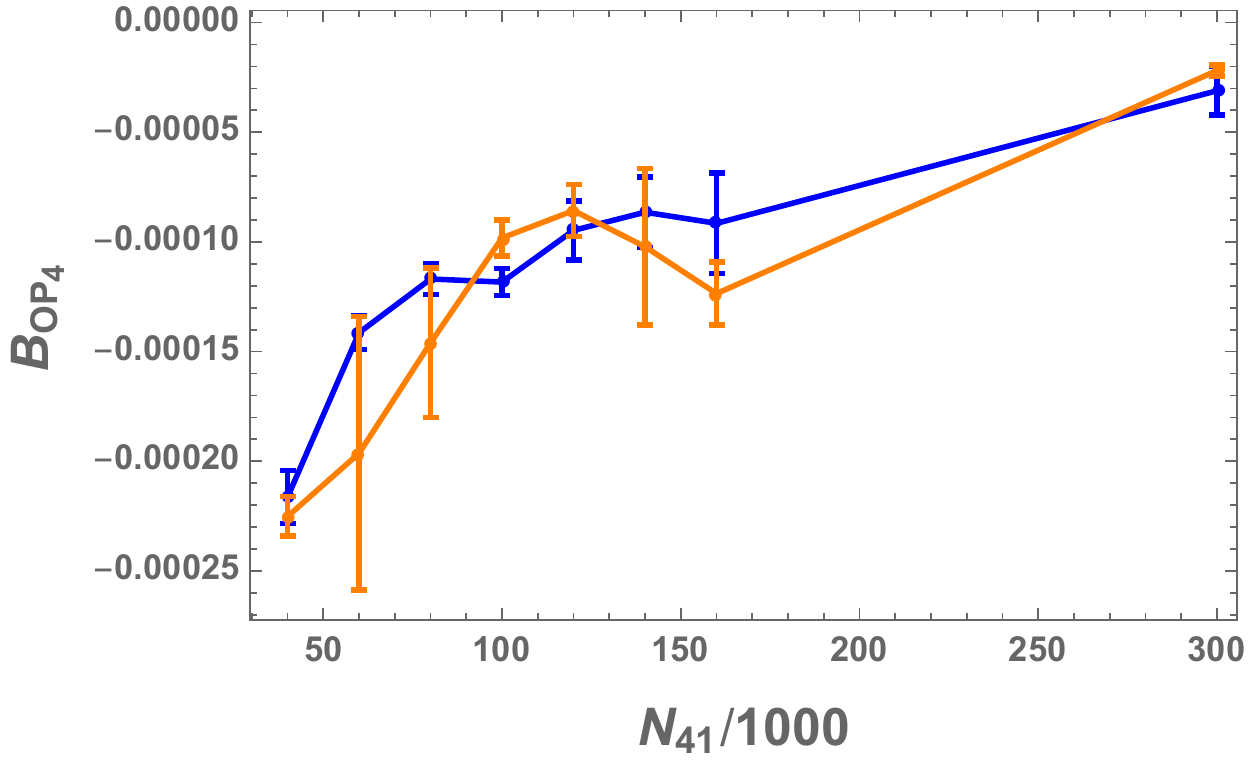}
\caption{\small Lattice volume dependence of the Binder cumulant \rf{binder} minima $B^{min}_{OP_{i}}(N_{4,1})$ ($i=1,...,4$) in CDT with toroidal spatial topology and for fixed $\kappa_0=2.2$. Blue data points are for the MC series started from a triangulation in phase $B$ while orange data points were started from a triangulation in phase $C_b$. {Error bars were estimated using a single-elimination (binned) jackknife procedure, where the bin sizes were selected in such a way that the statistical errors are maximized.}}
\label{N41bin}
\end{figure}

\end{section}

\begin{section}{Summary and conclusions}\label{summary}

Applying  phase transition analysis methods described in Section \ref{method} to the $B-C_B$ transition in CDT with the toroidal spatial topology we have shown that the transition is most likely the higher-order phase transition. This result is supported both by the finite size scaling analysis of equation \rf{scaling} showing the best fit scaling exponent $\nu = 2.7  > 1$, by the large volume behaviour of the Binder cumulant minima \rf{Bmin}: $B_{OP_i}^{min}(N_{4,1}\to \infty)\to 0$  and by the lack of hysteresis / two-state jumping of the order parameters measured at the (pseudo) critical points. 

The above result and also numerical values of the critical scaling exponent $\nu=2.7\pm0.4$ and the true critical point $\Delta^c(\infty)=0.073\pm0.004$ are also consistent with the $B-C_b$ transition measured in CDT with the spherical spatial topology for the same fixed value of the $\kappa_0=2.2$  parameter, where $\Delta^c(\infty)=0.077\pm0.004$ and  $\nu=2.51\pm0.03$, respectively \cite{Ambjorn:2012ij}. Thus the $B-C_b$ transition properties are the same in both  spatial topologies. This is also the case for the $A-C$  transition which was found to be the first-order phase transition in both topologies - the detailed analysis of the $A-C$ transition in the spherical and the toroidal CDT for various  Monte Carlo simulations' parameters (lattice volume fixing methods and  lengths of the (integer) time period~$T$) can be found in \cite{Ambjorn:2019pkp}. One can therefore formulate a conjecture that CDT results including the phase structure and the order of phase transitions are independent of the spatial topology choice, which is a parameter put in "by hand". 

\begin{figure}[H]
\centering
\includegraphics[width=0.55\linewidth]{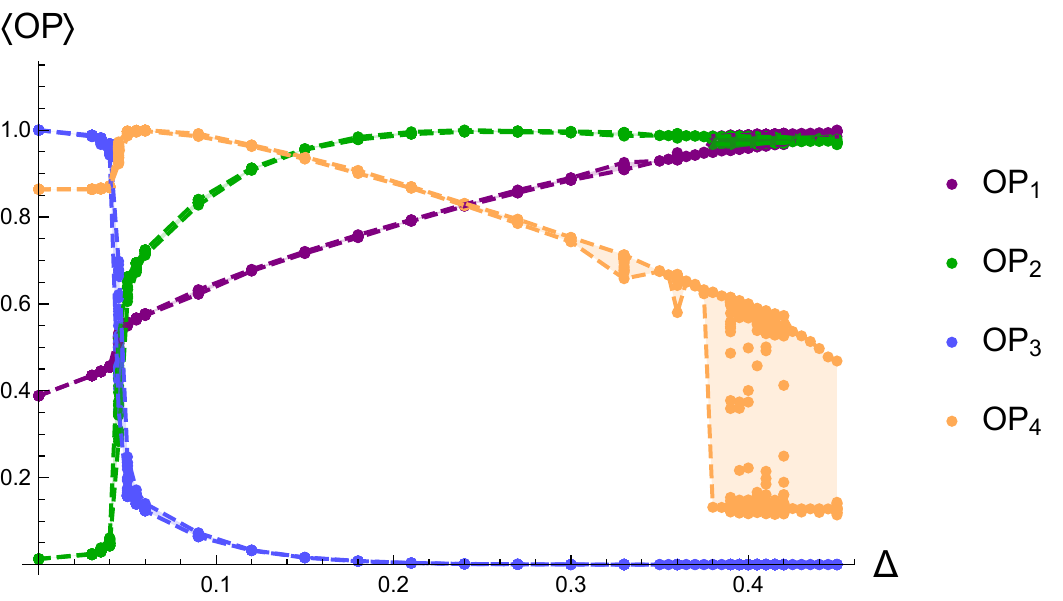}
\caption{\small Rescaled order parameters $\langle OP_1 \rangle, ..., \langle OP_4\rangle$ in CDT with the toroidal spatial topology measured   { for the (target) lattice volume $\bar N_{4,1}= 120k$ and $T=4$ time slices. Data were measured} for many different starting triangulations for each $\Delta$ ($\kappa_0 = 2.0$ is kept fixed), the number of starting configurations being different for various $\Delta$. Each data point denotes $\langle OP_i \rangle$ ($i=1,2,3,4$) measured from last $100\mathrm{k}$ sweeps (1~sweep = $10^7$ attempted MC moves), data from initial thermalization period were skipped. Shaded regions between the dashed lines denote the range of the measured data. Hysteresis is clearly visible for $\Delta \geq 0.38$, especially for the $OP_4$ parameter which is the most sensitive to the $C-C_b$   transition. This is not the case for the higher-order $B-C_b$ transition (described herein) observed around $\Delta\approx 0.05$.}
\label{hysteresis}
\end{figure}

The question mark remains for the $C-C_b$ transition which was found to be the higher-order phase transition in the spherical CDT \cite{Coumbe:2015oaa,Ambjorn:2017tnl}\footnote{Recent studies based on spectral properties of three-dimensional time slices in the spherical CDT  \cite{Clemente:2018,Clemente:2019} also indicate that the $C-C_b$ transition is  most likely the higher-order phase transition.} and has not been yet  investigated in detail in the toroidal CDT. The reason is that in the toroidal CDT case  one observes a  very strong hysteresis in the $C-C_b$ transition region\footnote{{The hysteresis is observed for sufficiently large (target) lattice volumes $\bar N_{4,1}$ such that the the three-volume of each (integer) time slice $\sim \bar N_{4,1}/T$ is big  enough  to allow for creation of high-order vertices, for small  $\bar N_{4,1}$ the bifurcation phase is not observed which is a finite-volume / discretization artifact.}} (see Figure \ref{hysteresis}) and therefore one is not able to perform precise MC measurements which would enable one to make finite size scaling analysis as it was explained in Section \ref{method}. The very strong hysteresis would suggest that the $C-C_b$ transition is most likely the first-order transition in the toroidal CDT, i.e. the order of the transition would  change due to the different spatial topology. But this can be as well an  algorithmic issue of the MC code used in the CDT simulations  and   more advanced methods should be used in order to resolve this problem.\footnote{We are currently working on a "multi-canonical" Monte Carlo algorithm which should enable one to measure both sides of the hysteresis in a single MC run and thus to estimate the lattice volume dependence of the hysteresis size and the position of pseudo-critical points  with much better precision.} 
In the toroidal CDT one was also able to make MC simulations in the most interesting region of the CDT parameter space, namely in the vicinity of the two "triple" points where the $A-B-C$ and the $B-C-C_b$ phases meet (see the CDT phase diagram in Figure \ref{pdnew}), which was not possible in the spherical CDT where MC simulations got effectively "frozen" in this region of the phase diagram. As a result in the toroidal CDT  one  observes the direct $B-C$ transition which was classified to be the first-order transition, albeit with some atypical properties suggesting a possible higher-order transition \cite{Ambjorn:2019lrm}. 
 Summing up, we have shown that the $B-C_b$ transition is the higher order transition which most likely makes the $B-C-C_b$ "triple" point the higher order transition point even though the $B-C$  and the  $C-C_b$ transitions are possibly the first-order transitions. The above "triple" point is thus a natural candidate for an UV fixed point for QG    \cite{Ambjorn:2014gsa,Ambjorn:2020}.
\end{section}

\section*{Acknowledgements}
JA acknowledges support from the Danish Research Council grant no. 7014-00066B {\it Quantum \linebreak Geometry}. 
JGS acknowledges support from the grant no. 2016/23/ST2/00289 from the National Science Centre, Poland.
AG  acknowledges 
support by the National Science Centre, Poland, under grant no. 2015/17/D/ST2/03479. 
JJ acknowledges support from the National Science Centre, Poland, 
grant no. 2019/33/B/ST2/00589.
DN acknowledges support from the National Science Centre, Poland, under grants no. 2015/17/D/ST2/03479 and 2019/32/T/ST2/00389.


\end{document}